\documentclass[fleqn,usenatbib]{mnras}
\usepackage[T1]{fontenc}
\usepackage{ae,aecompl}
\usepackage{graphicx}	
\usepackage{amsmath}	
\usepackage{amssymb}	
\usepackage{booktabs}
\usepackage{hyperref}
\usepackage{bm}

\newcommand{\cms}{\text{cm~s$^{-1}$}}
\newcommand{\kms}{\text{km~s$^{-1}$}}

\newcommand{\gpercmcu}{\text{g~cm$^{-3}$}}
\newcommand{\msun}{\text{M$_\odot$}}
\newcommand{\msunperyr}{\text{M$_\odot$~yr$^{-1}$}}

\defcitealias{2017MNRAS.468.3850S}{Paper I}

\title[Ring and gap formation]{The formation of rings and gaps in magnetically coupled disk-wind systems: ambipolar diffusion and reconnection}

\author[Suriano et al.]{
Scott S. Suriano$^{1}$\thanks{E-mail: suriano@virginia.edu},
Zhi-Yun Li$^{1}$, Ruben Krasnopolsky$^{2}$, and Hsien Shang$^{2}$
\\
$^{1}$Department of Astronomy, University of Virginia, Charlottesville, VA 22904, USA \\
$^{2}$Academia Sinica, Institute of Astronomy and Astrophysics, Taipei 10617, Taiwan\\
}

\date{Accepted XXX. Received YYY; in original form ZZZ}

\pubyear{2017}

\begin{document}
\label{firstpage}
\pagerange{\pageref{firstpage}--\pageref{lastpage}}
\maketitle

\begin{abstract}
Radial substructures in circumstellar disks are now routinely observed by ALMA. There is also growing evidence that disk winds drive accretion in such disks. We show through 2D (axisymmetric) simulations that rings and gaps develop naturally in magnetically coupled disk-wind systems on the scale of tens of au, where ambipolar diffusion (AD) is the dominant non-ideal MHD effect. In simulations where the magnetic field and matter are moderately coupled, the disk remains relatively laminar with the radial electric current steepened by AD into a thin layer near the midplane. The toroidal magnetic field sharply reverses polarity in this layer, generating a large magnetic torque that drives fast accretion, which drags the poloidal field into a highly pinched radial configuration. The reconnection of this pinched field creates magnetic loops where the net poloidal magnetic flux (and thus the accretion rate) is reduced, yielding dense rings. Neighbouring regions with stronger poloidal magnetic fields accrete faster, forming gaps. In better magnetically coupled simulations, the so-called `avalanche accretion streams' develop continuously near the disk surface, rendering the disk-wind system more chaotic. Nevertheless, prominent rings and gaps are still produced, at least in part, by reconnection, which again enables the segregation of the poloidal field and the disk material similar to the more diffusive disks. However, the reconnection is now driven by the non-linear growth of MRI channel flows. The formation of rings and gaps in rapidly accreting yet laminar disks has interesting implications for dust settling and trapping, grain growth, and planet formation.
\end{abstract}

\begin{keywords}
accretion, accretion discs -- magnetohydrodynamics (MHD) -- ISM: jets and outflows -- protoplanetary discs
\end{keywords}

\section{Introduction}
Observations with the Atacama Large Millimeter/submillimeter Array (ALMA) have revealed that many circumstellar disks contain intricate radial substructures \citep{2015ApJ...808L...3A,2016ApJ...820L..40A,2016ApJ...818L..16Z,2016ApJ...819L...7N,2016Sci...353.1519P,2016PhRvL.117y1101I,2016Natur.535..258C,2017A&A...597A..32V,2017A&A...600A..72F}, opening the door for previously inaccessible studies of the physical nature of disks. A number of physical processes have been proposed to explain the formation of rings and gaps in disks, including gap clearing by planets \citep{2015ApJ...809...93D,2015MNRAS.453L..73D,2017ApJ...843..127D,2017ApJ...850..201B}, rapid pebble growth at the condensation fronts of abundant volatile species \citep{2015ApJ...806L...7Z}, the pileup of volatile ices in sintering zones just outside snow lines \citep{2016ApJ...821...82O}, sharp changes in the disk viscosity at the boundaries of non-turbulent `dead zones' \citep{2015A&A...574A..68F,2016A&A...590A..17R}, magnetic self-organization through zonal flows \citep{2017A&A...600A..75B}, and the secular gravitational instability \citep{2014ApJ...794...55T}.

We presented, in \citet{2017MNRAS.468.3850S} (herein, \citetalias{2017MNRAS.468.3850S}), a novel mechanism for forming rings and gaps in magnetically coupled disk-wind systems in the presence of Ohmic resistivity, which is the dominant non-ideal magnetohydrodynamic (MHD) effect in the inner (sub-au) part of the disk \citep{2007Ap&SS.311...35W,2014prpl.conf..411T}. It relies on a magnetically driven disk wind \citep{1982MNRAS.199..883B} to remove angular momentum from the disk at a rate that varies strongly with radius, leading to a large spatial variation in accretion rate and thus the disk surface density. Observationally, there is now growing evidence for  rotating winds removing angular momentum from disks \citep{2016ApJ...831..169S,2017NatAs...1E.146H,2017A&A...607L...6T,2018arXiv180203668L}. Theoretically, a picture of wind-driven disk evolution is also beginning to emerge, with non-ideal MHD effects (Ohmic resistivity, ambipolar diffusion, and the Hall effect) suppressing MHD turbulence from the magnetorotational instability (MRI; \citealt{1991ApJ...376..214B}) over a wide range of radii, which leaves MHD disk winds as the primary driver of disk accretion in these regions \citep{2000ApJ...530..464F,2003ApJ...585..908F,2011ApJ...736..144B,2013ApJ...769...76B,2013MNRAS.434.2295K,2015ApJ...801...84G}.

In this follow-up work, we focus on the intermediate radii of young star disks (a few to tens of au) where ambipolar diffusion (AD) starts to become the most important non-ideal MHD effect, especially in the upper layers of the disk \citep{2007Ap&SS.311...35W,2014prpl.conf..411T}. We find that rings and gaps are naturally produced in the presence of a significant poloidal magnetic field, just as in the resistive case studied in \citetalias{2017MNRAS.468.3850S}. We show that a relatively laminar disk-wind system develops in the presence of a relatively strong ambipolar diffusion, which makes it easier to analyse the simulation results and identify a new mechanism for ring and gap formation. The mechanism is driven by reconnection of the highly pinched poloidal magnetic field in a midplane current sheet steepened by ambipolar diffusion \citep{1994ApJ...427L..91B}, in a manner that is reminiscent of the tearing mode \citep{1963PhFl....6..459F} or the pinch-tearing mode \citep{2009MNRAS.394..715L}. We show that the reconnection leads to weakening of the poloidal field in some regions, which accrete more slowly and form rings, and field concentration in others, where accretion is more efficient creating gaps.

The rest of the paper is organized as follows. In Section~\ref{sec:setup} we describe the simulation setup, including the equations solved, the initial disk model, and the boundary conditions. Section~\ref{sec:ref} analyses the results of a reference simulation in detail and explains how rings and gaps are formed in the coupled disk-wind system in the presence of relatively strong ambipolar diffusion. In Section~\ref{sec:param} we explore how changes in the magnetic field and ambipolar diffusion strength modify the picture of the reference run. In Section~\ref{sec:discuss} we compare to other similar works in the field and discuss the implications of our work on dust settling, growth, and trapping that are important to the formation of planetesimals and planets. Finally, Section~\ref{sec:conc} concludes with the main results of this study.

\section{Problem setup}\label{sec:setup}

\subsection{MHD equations}
We use the ZeusTW code \citep{2010ApJ...716.1541K} to solve the time-dependent magnetohydrodynamic (MHD) equations in axisymmetric spherical coordinates ($r,\theta,\phi$). The ZeusTW code is based on the ideal MHD code, ZEUS-3D (version 3.4; \citealt{1996ApJ...457..291C,2010ApJS..187..119C}), which is itself developed from ZEUS-2D \citep{1992ApJS...80..753S,1992ApJS...80..791S}. In the ZeusTW code, Ohmic resistivity is treated using the algorithm described in \citet{2000ApJ...530..464F} and AD is implemented using the fully explicit method of \citet{1995ApJ...442..726M} (see also \citealt{2011ApJ...738..180L}). The equations solved are
\begin{equation}
\frac{\partial \rho}{\partial t} + \nabla \cdot \left( \rho \bm{v} \right) = 0, 
\end{equation}
\begin{equation}
\rho\frac{\partial\bm{v}}{\partial t} + \rho\left(\bm{v}\cdot\nabla\right)\bm{v} = -\nabla P + \bm{J}\times\bm{B}/c - \rho\nabla\Phi_g,
\end{equation}
\begin{equation}
\frac{\partial\bm{B}}{\partial t} = \nabla\times\left(\bm{v}\times\bm{B}\right) - \frac{4\pi}{c}\nabla\times\left(\eta_O\bm{J} + \eta_A\bm{J_\perp}\right),\label{eq:induction}
\end{equation}
\begin{equation}
\frac{\partial e}{\partial t} + \nabla \cdot \left(e \bm{v} \right) = -P \nabla \cdot \bm{v},
\end{equation}
where the internal energy is $e=P/(\Gamma-1)$ and $\Gamma$ is the adiabatic index. The current density is $\bm{J}=(c/4\pi)\nabla\times\bm{B}$ and the current density perpendicular to the magnetic field is $\bm{J}_\perp=-\bm{J}\times\bm{B}\times\bm{B}/B^2$. The Ohmic resistivity is $\eta_O$ and the effective ambipolar diffusivity $\eta_A$ is defined as
\begin{equation}
\eta_A=\frac{B^2}{4\pi\gamma\rho\rho_i},
\label{ADdiffusion}
\end{equation}
where $\rho_i$ is the ion density and $\gamma=\langle\sigma v\rangle_i/(m+m_i)$ is the frictional drag coefficient with units of $\rm{cm}^3~\rm{g}^{-1}~\rm{s}^{-1}$. The remaining parameters have their usual definitions. When referring to cylindrical coordinates, we will use the notation ($R,\phi,z$) such that $R=r\sin{\theta}$ and $z=r\cos{\theta}$.

\subsection{Initial conditions}
The initial conditions are similar to those in \citetalias{2017MNRAS.468.3850S}. We describe them here in detail for completeness. Specifically, the simulation domain is separated into two regions: a thin, cold, rotating disk orbiting a 1~\msun~central source at the grid origin and an initially non-rotating, hot corona above the disk that is quickly replaced by a magnetic wind driven from the disk. We choose the adiabatic index to be $\Gamma=1.01$ so that the material in the simulation domain is locally isothermal in the sense that any parcel of disk material nearly retains its initial temperature no matter where it moves. The initial temperature distribution is assumed to decrease with radius as a power-law $T\propto r^{-1}$, so that the sound speed is proportional to the local Keplerian speed.

\subsubsection{Disk}
The geometrically thin disk is characterized by the dimensionless parameter $\epsilon = h/r = c_s/v_K \ll 1$, where $h$ is the disk scale height, $c_s$ is the isothermal sound speed, and $v_K$ is the Keplerian speed. The initial value of $\epsilon$ is set to 0.05 for all simulations in this work. The disk is limited to the equatorial region where the polar angle $\theta \in {[\pi/2 - \theta_0,\pi/2 + \theta_0]}$, with disk (half) opening angle set to $\theta_0=\arctan(2\epsilon)$, i.e., the initial disk half-thickness is set to twice the scale height. This choice is somewhat arbitrary, but a more elaborate treatment of the initial disk surface is not warranted because the structure of the disk surface is quickly modified by a magnetic wind. The disk density takes the form of a radial power law multiplied by a Gaussian function of $z/r=\cos\theta$,
\begin{equation}
\rho_d(r,\theta) = \rho_{0} \left(\frac{r}{r_0}\right)^{-\alpha} \exp \left(-\frac{\cos^2\theta}{2 \epsilon^2}\right),
\end{equation}
as determined by hydrostatic balance. The subscript `0' refers to values on the disk midplane at the inner radial boundary. For all simulations shown in this paper, we use $\alpha=3/2$. The disk pressure is set as
\begin{equation}
P_d(r,\theta)=\rho_d c_s^2,
\end{equation}
with $c_s = \epsilon v_K$. The radial pressure gradient causes the equilibrium rotational velocity $v_\phi$ to be slightly sub-Keplerian,
\begin{equation}
v_\phi =  v_K \sqrt{1-(1+\alpha)\epsilon^2}.
\end{equation}

\subsubsection{Corona}
We require that the hydrostatic corona is initially in pressure balance with the disk surface. This constraint sets the density drop from the disk surface to the corona by $1/[(1+\alpha)\epsilon^2]=160$, and a corresponding increase in temperature from the disk surface to the corona by the same factor. Therefore, the coronal density and pressure are
\begin{equation}
\rho_{c}(r)=\rho_{0} \epsilon^2(1+\alpha)\exp\left[-\frac{\cos^2\theta_0}{2 \epsilon^2}\right] \left(\frac{r}{r_0}\right)^{-\alpha}\equiv\rho_{c,0}\left(\frac{r}{r_0}\right)^{-\alpha},
\end{equation}
\begin{equation}
P_c(r)= \rho_c v_K^2/(1+\alpha).
\end{equation}
It is, however, important to note that the initial hot coronal material is quickly replaced by the colder disk material that remains nearly isothermal as it is launched from the disk surface into a wind.

\subsubsection{Magnetic field}
To ensure that the magnetic field is divergence-free initially, we set the magnetic field components using the magnetic flux function $\Psi$ as in \citet{2007A&A...469..811Z},
\begin{equation}
\Psi(r,\theta) = \frac{4}{3}r_0^2 B_{\mathrm{p},0}\left(\frac{r\sin\theta}{r_0}\right)^{3/4} \frac{m^{5/4}}{\left(m^{2}+\cot^2\theta\right)^{5/8}}\label{eq:psi},
\end{equation}
where $B_{\mathrm{p},0}$ sets the scale for the poloidal field strength and the parameter $m$ controls the bending of the field. The value of $B_{\mathrm{p},0}$ is set by the initial plasma-$\beta$, the ratio of the thermal to magnetic pressure, on the disk midplane, which is approximately $10^3$ ($0.922\times 10^3$ to be more exact) for most of the simulations. Since varying $m$ from 0.1 to 1 has little effect on the long-term disk or wind magnetic field structure \citep{2014ApJ...793...31S}, we use $m=0.5$ for all simulations presented in this work. The initial magnetic field components are then calculated as
\begin{equation}
B_r=\frac{1}{r^2\sin{\theta}}\frac{\partial\Psi}{\partial\theta},
\end{equation}
\begin{equation}
B_\theta=-\frac{1}{r\sin\theta}\frac{\partial\Psi}{\partial r}.
\end{equation}

\subsection{Ambipolar diffusion}\label{sec:AD}
The magnetic diffusivities associated with non-ideal MHD effects, including ambipolar diffusion, depend on the densities of charged particles, which can in principle be computed through detailed chemical networks (e.g., \citealt{2009ApJ...701..737B}). Here, as a first step toward a comprehensive model, we will simply parametrize the density of ions as
\begin{equation}
\rho_i = \rho_{i,0} f(\theta) \left(\frac{\rho}{\rho_0}\right)^{\alpha_{AD}},
\end{equation}
where
\begin{equation}\label{eq:ftheta}
  f(\theta) =
  \begin{cases}
    \exp\left(\frac{\cos^2(\theta+\theta_0)}{2 \epsilon^2}\right) & \theta<\pi/2-\theta_0 \\
    1                                                             & \pi/2-\theta_0\leq\theta\leq\pi/2+\theta_0 \\
    \exp\left(\frac{\cos^2(\theta-\theta_0)}{2 \epsilon^2}\right) & \theta>\pi/2+\theta_0.
  \end{cases}
\end{equation}
The angular dependence $f(\theta)$ is chosen such that, at a given radius, the ion density increases rapidly in the tenuous disk atmosphere, to mimic the ionization by high energy photons (UV and X-rays) from the central young star in addition to cosmic rays (e.g., \citealt{1981PASJ...33..617U,2011ApJ...735....8P,2017MNRAS.472.2447G}). In the simulations presented in this work, we take $\alpha_{AD}=0.5$. This power-law dependence for the ion density is roughly what is expected when the volumetric cosmic ray ionization rate is balanced by the recombination rate of ions and electrons, under the constraint of charge neutrality (i.e., $\zeta n\propto n_e n_i \propto n_i^2$, where $\zeta$ is the cosmic ray ionization rate per hydrogen nucleus; see page 362 of \citealt{1992phas.book.....S}). 

The magnitude of the ion density, and therefore the ion-neutral drag force, $\bm{F_d}=\gamma\rho\rho_i(\bm{v_i}-\bm{v})$, is sometimes quantified through the dimensionless ambipolar Elsasser number \citep{2007NatPh...3..604C,2011ApJ...727....2P,2011ApJ...736..144B},
\begin{equation}
\Lambda=\frac{\gamma\rho_i}{\Omega}=\frac{v_A^2}{\eta_A\Omega},
\end{equation}
where $\gamma$ is the frictional drag coefficient. Physically, the Elsasser number is the collision frequency of a neutral particle in a sea of ions of density $\rho_i$, normalized to the Keplerian orbital frequency. For example, the Elsasser number will be unity when the neutral particle collides $2\pi$ times in one orbital period. As the neutral-ion collision frequency increases to infinity, so does the Elsasser number, and the bulk neutral medium becomes perfectly coupled to the ions/magnetic field (i.e., the ideal MHD limit). Similarly, as the Elsasser number drops to zero, the neutrals and ions no longer collide; the neutrals are entirely decoupled from the magnetic field. For our reference simulation, we choose the Elsasser number to be $\Lambda_0=0.25$ at the inner boundary on the disk midplane, but will vary this parameter to gauge its effects on the coupled disk-wind system. The choice of $\alpha_{AD}=0.5$, assuming that the drag coefficient $\gamma$ is constant, implies that the Elsasser number is proportional to $r^{3/4}$, thus larger radii are better coupled than smaller radii in the reference simulation.

In some cases, we have considered an explicit Ohmic resistivity $\eta_O$ in addition to ambipolar diffusion. In these cases, it is useful to compute the effective ambipolar diffusivity $\eta_A$ according to equation~(\ref{ADdiffusion}), to facilitate the comparison of the two effects. 

\subsection{Grid}\label{sec:grid}
The equations are solved for $r\in{[1,100]}$~au and $\theta\in{[0,\pi]}$. The radial grid limits are chosen such that they encompass the anticipated disk region where ambipolar diffusion is the most important non-ideal effect, especially in the upper layers of the disk \citep{2014prpl.conf..411T}. A `ratioed' grid is used in the radial direction such that $dr_{i+1}/dr_i=1.012$ is constant and $r_{i+1}=r_i+dr_i$. The grid spacing at the inner edge is set as $dr_0=2.3r_0d\theta$. The grid is uniform in $\theta$ with a resolution of $n_r\times n_\theta = 400\times720$. This results in 24 grid cells from the disk midplane to surface (at two disk scale heights) in the reference run.

\subsection{Boundary conditions}\label{sec:bc}

Both the inner and outer radial boundaries use the standard outflow condition in Zeus codes, where the flow quantities in the first active zone are copied into the ghost zones except for the radial component of the velocity, $v_r$, which is set to zero in the ghost zones if it points into the computation domain in the first active zone (i.e., if $v_r > 0$ in the first active zone at the inner radial boundary or $v_r < 0$ in the first active zone at outer radial boundary). The standard axial reflection boundary condition is used on the polar axis ($\theta=0$ and $\pi$), where the density and radial components of the velocity and magnetic field ($v_r$ and $B_r$) in the ghost zones take their values in the corresponding active zones while the polar and azimuthal components ($v_\theta$, $B_\theta$, and $v_\phi$) take the negative of their values in the corresponding active zones. We set $B_\phi$ to vanish on the polar axis and on the inner radial boundary, since it is taken to be non-rotating.

\section{Reference model and the formation of rings and gaps through reconnection}\label{sec:ref}
We will first discuss in detail a reference simulation to which other simulations with different parameters can be compared. The parameters used in all simulations are listed in Table~\ref{tab:sims}. The initial density at $r_0$ on the disk midplane is $\rho_0=1.3\times 10^{-10}~\gpercmcu$. Of the parameters to be changed in later simulations, this `reference' simulation uses $\beta_0\sim10^3$ and $\Lambda_0=0.25$. Note that the radial power-law dependence of the ambipolar Elsasser number is $\Lambda\propto r^{3/4}$, whereas the initial distribution of plasma-$\beta$ is constant with radius.

\begin{figure*}
\centering
\includegraphics[width=2.0\columnwidth]{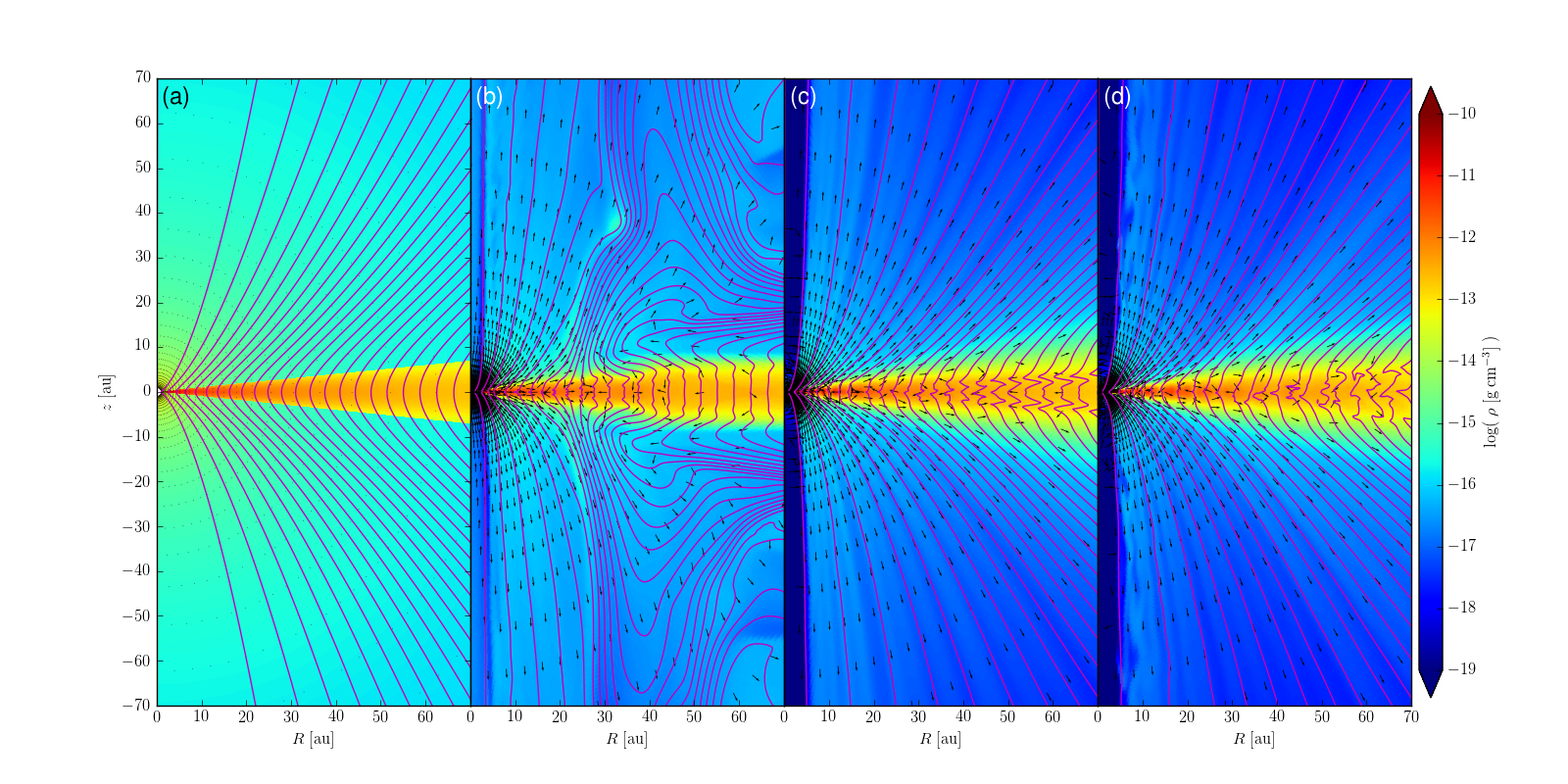}
\caption{A representative (`reference') axisymmetric simulation. Shown is the mass volume density (logarithmically spaced colour contours in units of \gpercmcu), the poloidal magnetic field lines (magenta), and the poloidal velocity unit vectors (black). Panels (a)-(d) corresponding to simulation times of 0, 200, 1000, and 2500 inner orbital periods, respectively. (See the supplementary material in the online journal for an animated version of this figure.)}
\label{fig:global}
\end{figure*}

\subsection{Global picture}
The overall global evolution of the disk is seen in Fig.~\ref{fig:global}. The frames show the simulation at times of 0, 200, 1000, and 2500$t_0$ (left to right), where $t_0=1$~yr is the orbital period at the inner edge of the simulation domain ($r_0=1$~au). The disk wind launching proceeds in an inside-out fashion, i.e, the wind is launched from larger disk radii as the simulation progresses. By 1000$t_0$ (one orbit at the outer radius $r=100$~au), the simulation appears to have no memory of the initial simulation setup and the magnetic field geometry is conducive to launching an MHD disk wind at all radii. From initial inspection, the disk wind that is launched is very steady in time (see the supplementary material online for an animated version of Fig.~\ref{fig:global}). The magnetic flux is also very much fixed in time following an initial adjustment period of approximately $t/t_0=100$. From this point on, the poloidal magnetic field lines stay in close proximity to their equilibrium footpoint in the disk (magenta lines in Fig.~\ref{fig:global}). However, there is persistent laminar accretion radially inward through the disk and across magnetic field lines.

\begin{figure}
\centering
\includegraphics[width=1.0\columnwidth]{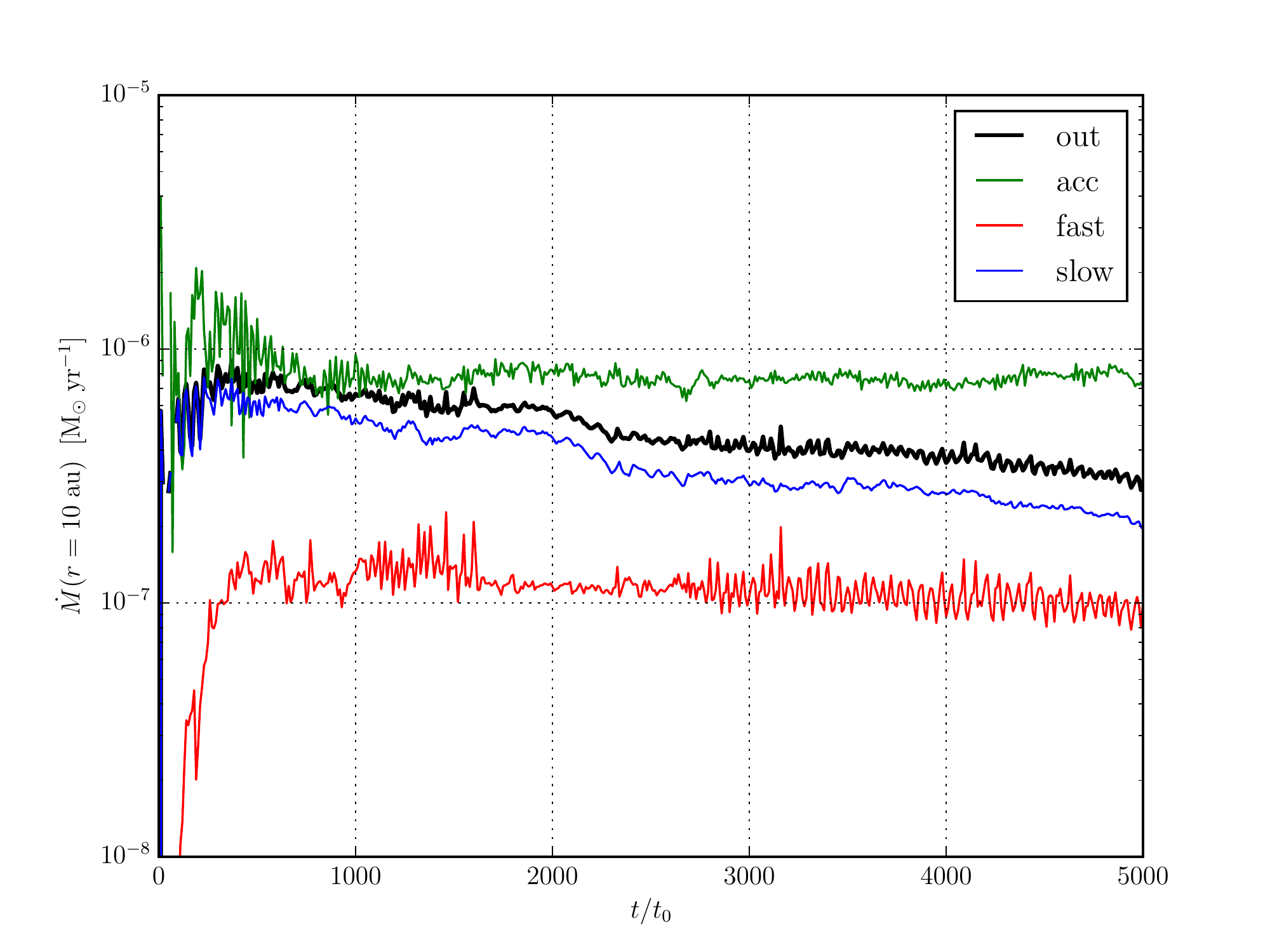}
\caption{The mass outflow rates (\msunperyr) as a function of time in the reference simulation through a sphere of radius $r=10$~au. The total mass outflow rate both above and below the disk ($\vert \pi/2-\theta \vert>2\epsilon$) is shown in black. It is separated into two velocity components, with the fast component ($v_r>10$~\kms) shown in red and the slow component ($v_r\leq10$~\kms) in blue. The green line shows the mass accretion rate through the disk ($\vert \pi/2-\theta \vert<2\epsilon$).}
\label{fig:mdot}
\end{figure}

A relatively massive and steady wind is launched from the disk surface for almost the entire simulation. Figure~\ref{fig:mdot} plots the mass outflow rate for the wind through a sphere of radius $r=10$~au, excluding the disk region ($\vert\pi/2-\theta\vert<2\epsilon$; black line). The mass outflow rate slowly decreases for times $t/t_0\gtrsim300$ from a value of $\dot{M}_\mathrm{w}\approx7\times10^{-7}~\msunperyr$ down to approximately $3\times10^{-7}~\msunperyr$ by $t/t_0=5000$. Most of the mass lost in the wind moves rather slowly, with a radial expansion speed less than the local Keplerian rotation speed at $r=10$~au for a one solar-mass star ($v_r<10$~\kms; blue line). This picture is reminiscent of \citetalias{2017MNRAS.468.3850S}, where a weak initial magnetic field (again, $\beta\sim10^3$) is able to drive slow and massive outflow. There is, however, a substantial mass loss from a relatively fast outflow ($v_r > 10$~\kms) as well (red line). Note that the mass infall rate in the low-density polar region is of the order $10^{-13}~\msunperyr$, which is much smaller than the total mass outflow rate. The mass accretion rate through the disk at this same radius ($r=10$~au) is of the same order as the total outflow rate (green line). It stays relatively constant at $\dot{M}_\mathrm{acc}\approx 8\times10^{-7}~\msunperyr$, even as the outflow rate decreases slowly at later times. 

\subsection{Disk-wind connection}
\begin{figure*}
\centering
\includegraphics[width=2.0\columnwidth]{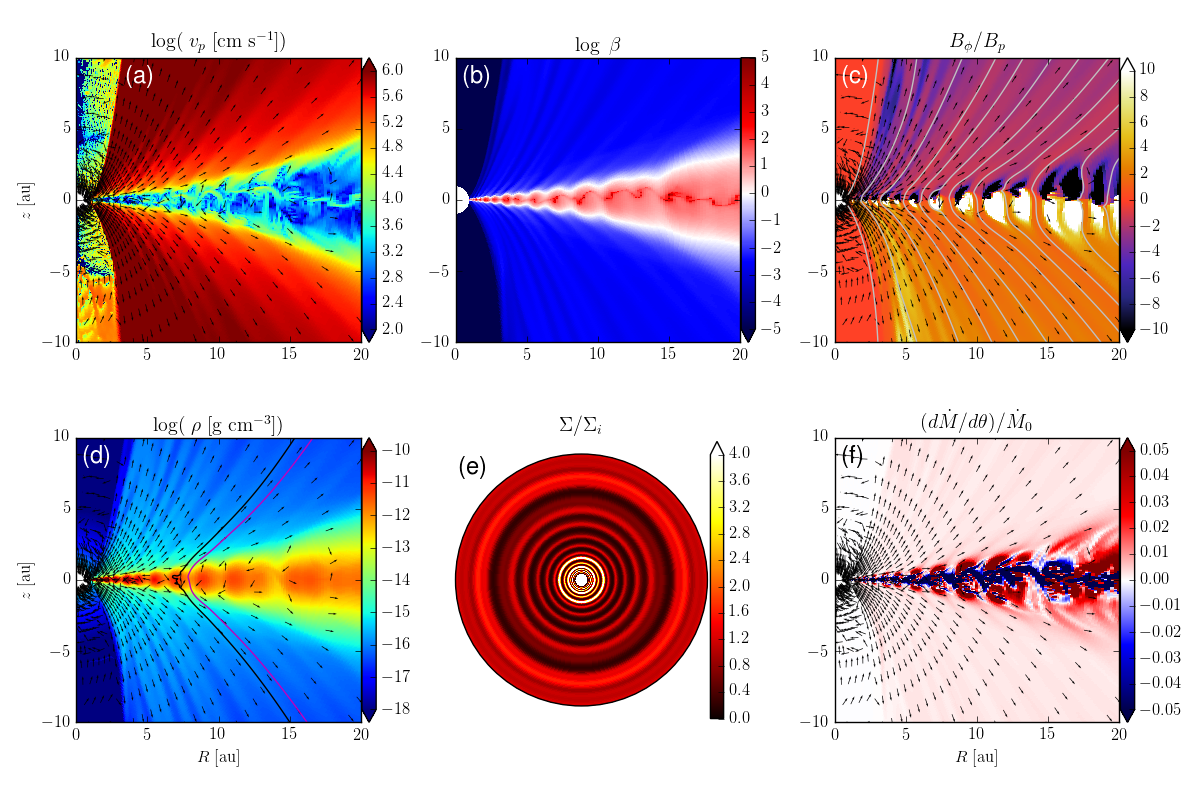}
\caption{The reference simulation at time $t/t_0=2500$. The colour contours show (a) the logarithm of poloidal velocity (\cms), (b) the logarithm of plasma-$\beta$, (c) the ratio of the toroidal to the poloidal magnetic field components, $B_\phi/B_\mathrm{p}$, (d) the logarithm of density (\gpercmcu), (e) an axisymmetric, face-on view of the disk surface density distribution normalized to the initial power-law distribution, $\Sigma_i=\Sigma_0(r/r_0)^{-1/2}$, and (f) the mass flux per unit polar angle, $d\dot{M}/d\theta=2\pi r^2\rho v_r\sin\theta$, normalized to $\dot{M}_0=r_0^2\rho_0 c_{s,0}$. Poloidal magnetic field lines (i.e., magnetic flux contours) are shown in grey in panel (c). Panel (d) shows two specific poloidal magnetic field lines with midplane footpoints at $r=8$~au (magneta; see Fig.~\ref{fig:path_8au}) and $r=7$~au (black; see Fig.~\ref{fig:path_7au}). Poloidal velocity unit vectors are plotted in black in panels (a), (c), (d), and (f). (See the supplementary material in the online journal for an animated version of this figure.)}
\label{fig:beta}
\end{figure*}

Figure~\ref{fig:beta} plots various physical quantities of the reference simulation at $t/t_0=2500$ up to a radius of $r=20$~au. Panel (a) shows the poloidal velocity as the accreting disk material is peeled off the disk surface and launched in a wind. The velocity of the disk accretion is of the order 1~m~s$^{-1}$, while in the wind, material is accelerated up to velocities of $v_\mathrm{p}\gtrsim 10$~\kms. The disk and wind regions are easily distinguishable in this plot. They are also quite distinct in panel (b), where the ratio of the thermal pressure to the magnetic pressure, i.e., plasma-$\beta$, is plotted. The initial $\beta$ at the disk midplane is $\sim 10^3$ and it is constant as a function of radius. At the time shown, the thermal and magnetic pressures become approximately equal at the base of the wind ($\beta\sim1$), while $\beta$ decreases to $10^{-2}$ or less in the wind zone. The bulk of the disk has plasma-$\beta$ of approximately 10, although there remains a thin layer where the plasma-$\beta$ peaks at a value slightly larger than the initial $\beta_0$. This thin layer is where the toroidal magnetic field goes to zero as it reverses direction; it is therefore a current layer (see also the simulations of \citealt{2013ApJ...769...76B,2013ApJ...772...96B,2015ApJ...801...84G,2017ApJ...836...46B,2017A&A...600A..75B,2017ApJ...845...75B}). This can be seen in panel (c), which plots the ratio of the toroidal magnetic field to the poloidal magnetic field, $B_\phi/B_\mathrm{p}$. The white regions are where $B_\phi>0$ in the disk and the black regions have $B_\phi<0$. The poloidal magnetic field lines (i.e., constant magnetic flux contours) are shown in grey. Note that the toroidal field greatly dominates the poloidal field in distinct radial locations, while the poloidal field is stronger (with a smaller $\vert B_\phi/B_\mathrm{p}\vert$) in adjacent regions. These regions with less twisted field lines correspond to regions with lower densities in the disk, as shown in panel (d). They will be referred to as `gaps.' The neighbouring regions, where the field lines are more twisted, have higher densities; they will be referred to as `rings.' The rings and gaps are shown more clearly in the face-on view of the disk in panel (e), where the distribution of the surface density (normalized to the initial distribution) is plotted. How such rings and gaps form will be discussed in detail in Section~\ref{sec:ring}. Here, we would like to point out that there is vigorous accretion (and some expansion) in both types of regions, as shown in panel (f), which plots the spatial distribution of the radial mass flux per unit polar angle, $d\dot{M}/d\theta=2\pi r^2\rho v_r\sin\theta$. The blue cells show negative mass flux or inward accretion and the red cells represent positive mass flux or outward expansion. We see that most of the accretion in the disk occurs along the thin current sheet previously mentioned. In disk regions above and below this layer, there is a strong variation in mass flux, with larger outward mass flux from rings compared to those from the adjacent gaps, although this strong variation does not appear to extend into the (faster) wind zone. 

\begin{figure*}
    \centering
    \includegraphics[width=1.0\columnwidth]{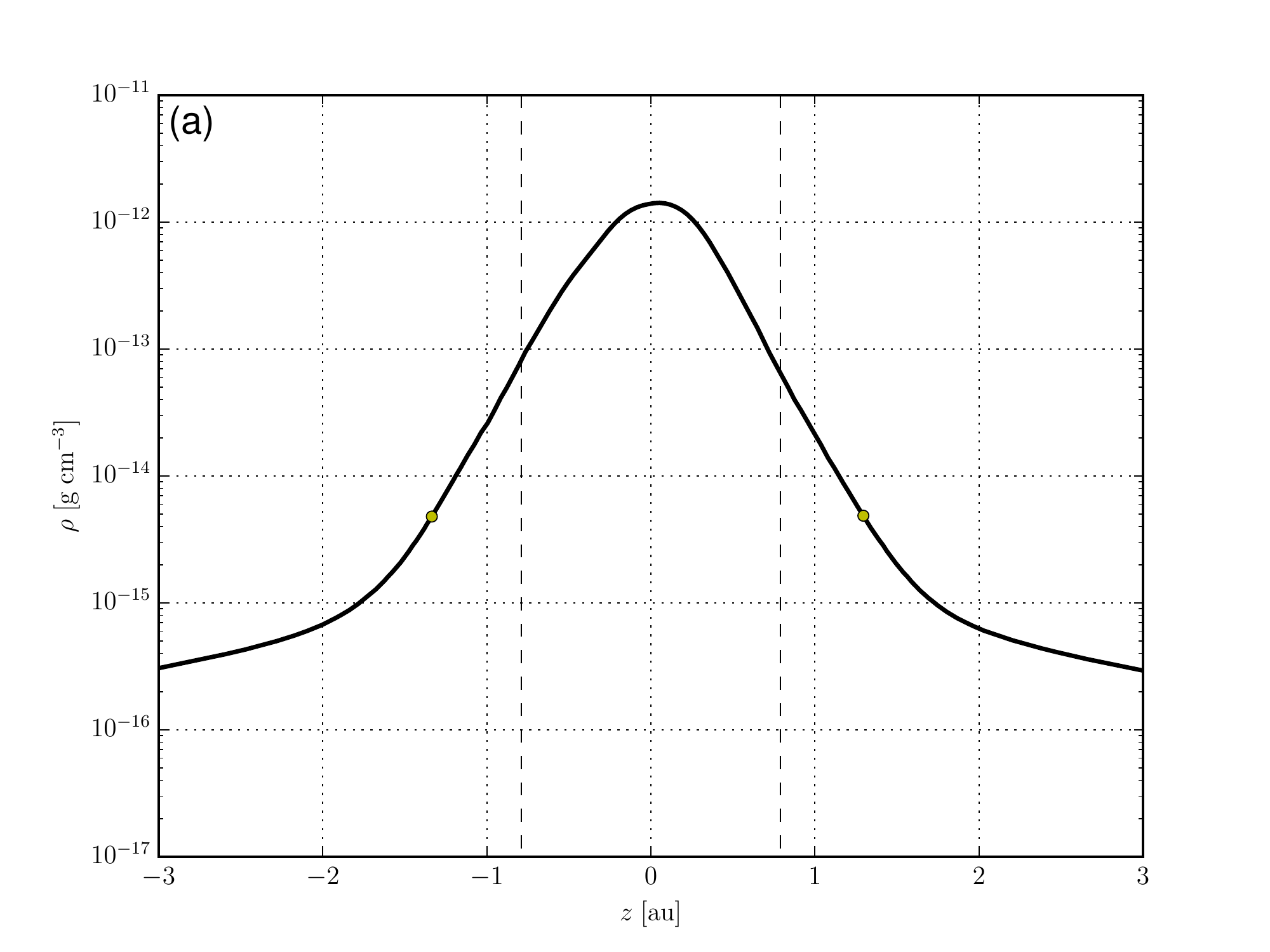}
    \includegraphics[width=1.0\columnwidth]{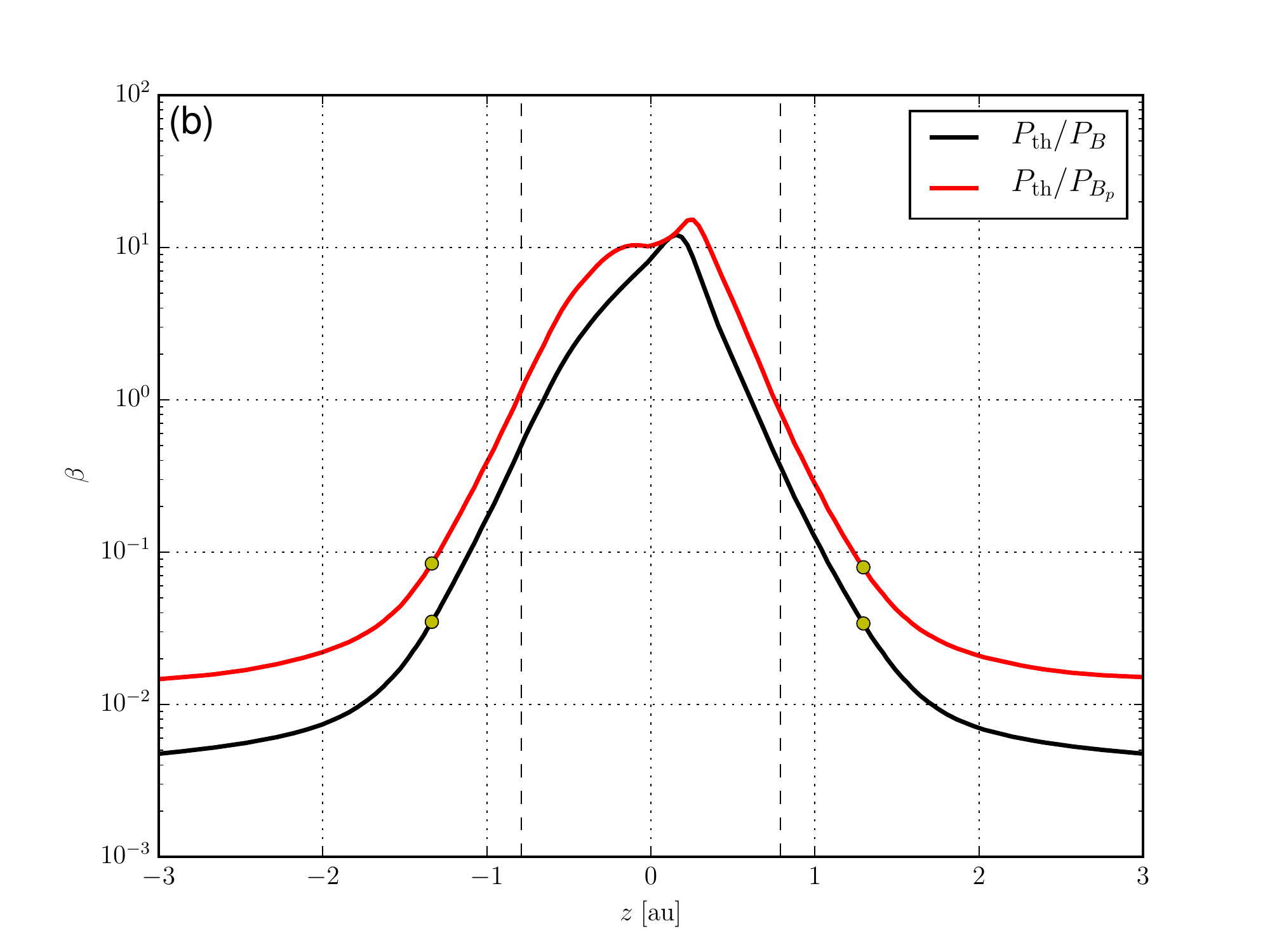}
    \includegraphics[width=1.0\columnwidth]{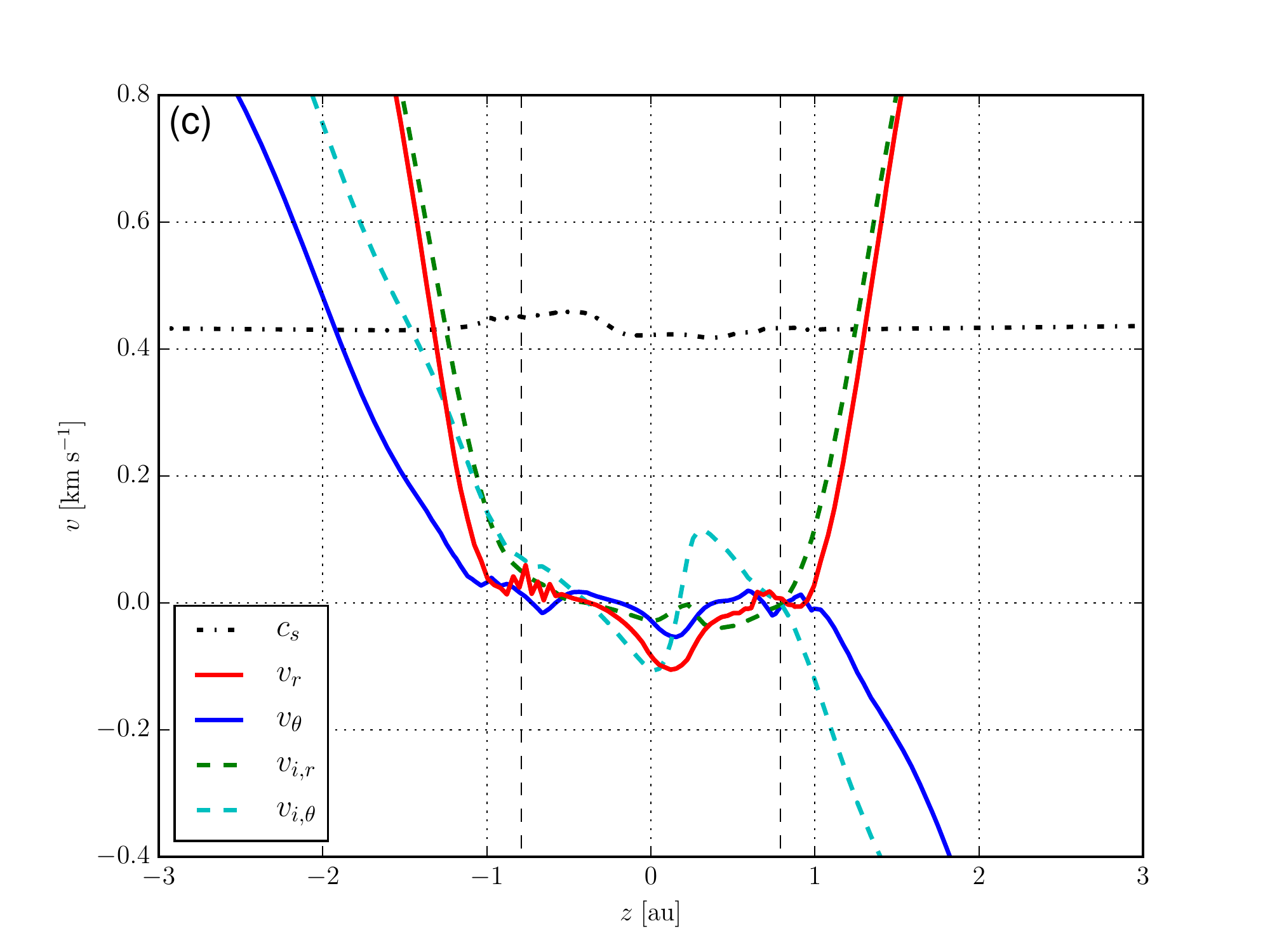}
    \includegraphics[width=1.0\columnwidth]{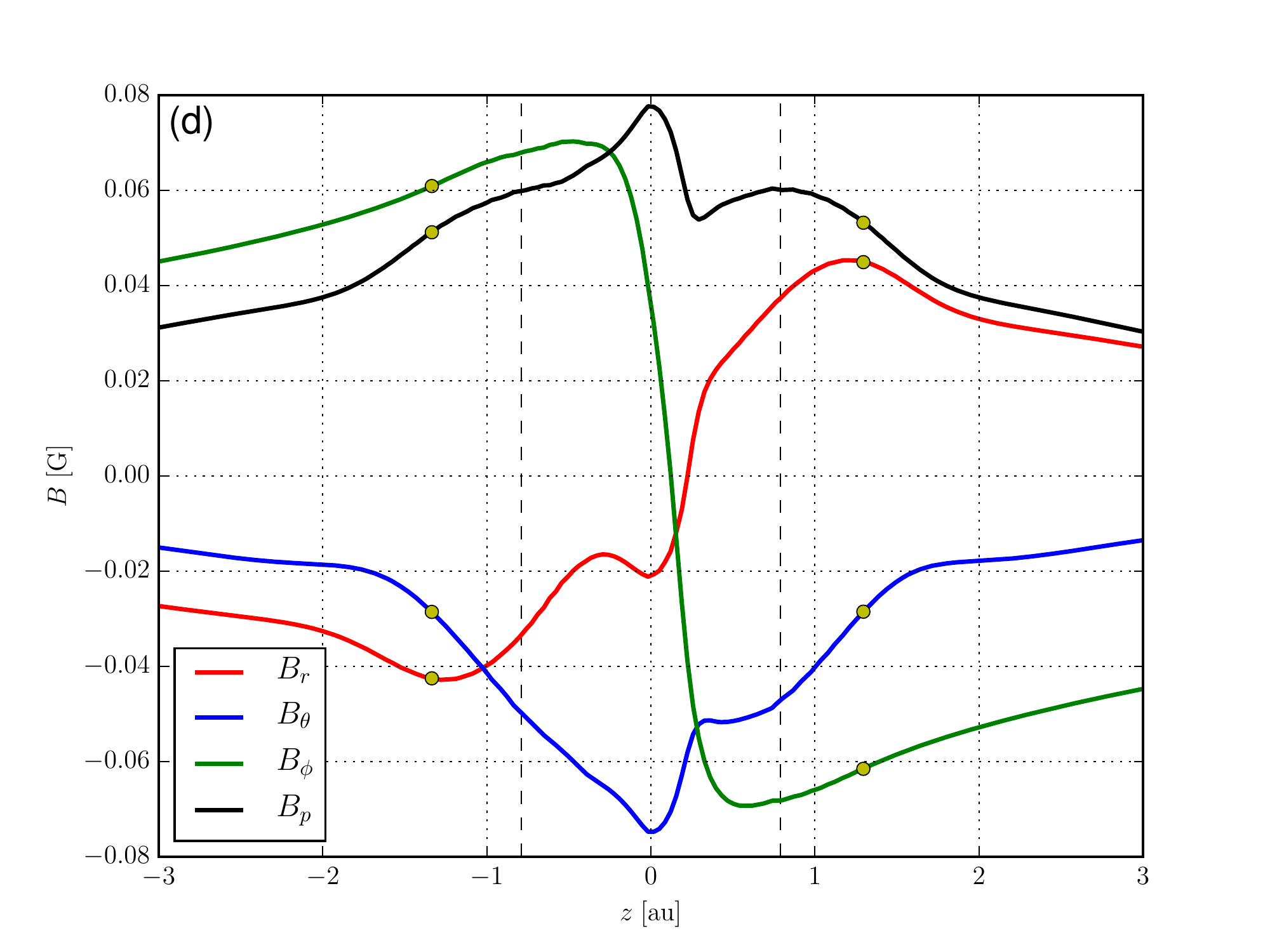}
    \caption{Physical quantities plotted along a poloidal magnetic field line as a function of the vertical height $z$. This representative field line passes through a low-density gap at $r=8$~au. The panels show (a) the density distribution, (b) plasma-$\beta$ for the total magnetic field strength (black) and for the poloidal magnetic field strength (red), (c) the poloidal components of the neutral (solid lines) and ion velocities (dashed lines), and the adiabatic sound speed (dash-dotted line), and (d) the magnetic field components. The yellow circles show the sonic point (where the poloidal velocity is equal to the adiabatic sound speed) and the vertical dashed lines show the initial disk height of $z=\pm2h_0$.}
    \label{fig:path_8au}
\end{figure*}

\begin{figure*}
    \centering
    \includegraphics[width=1.0\columnwidth]{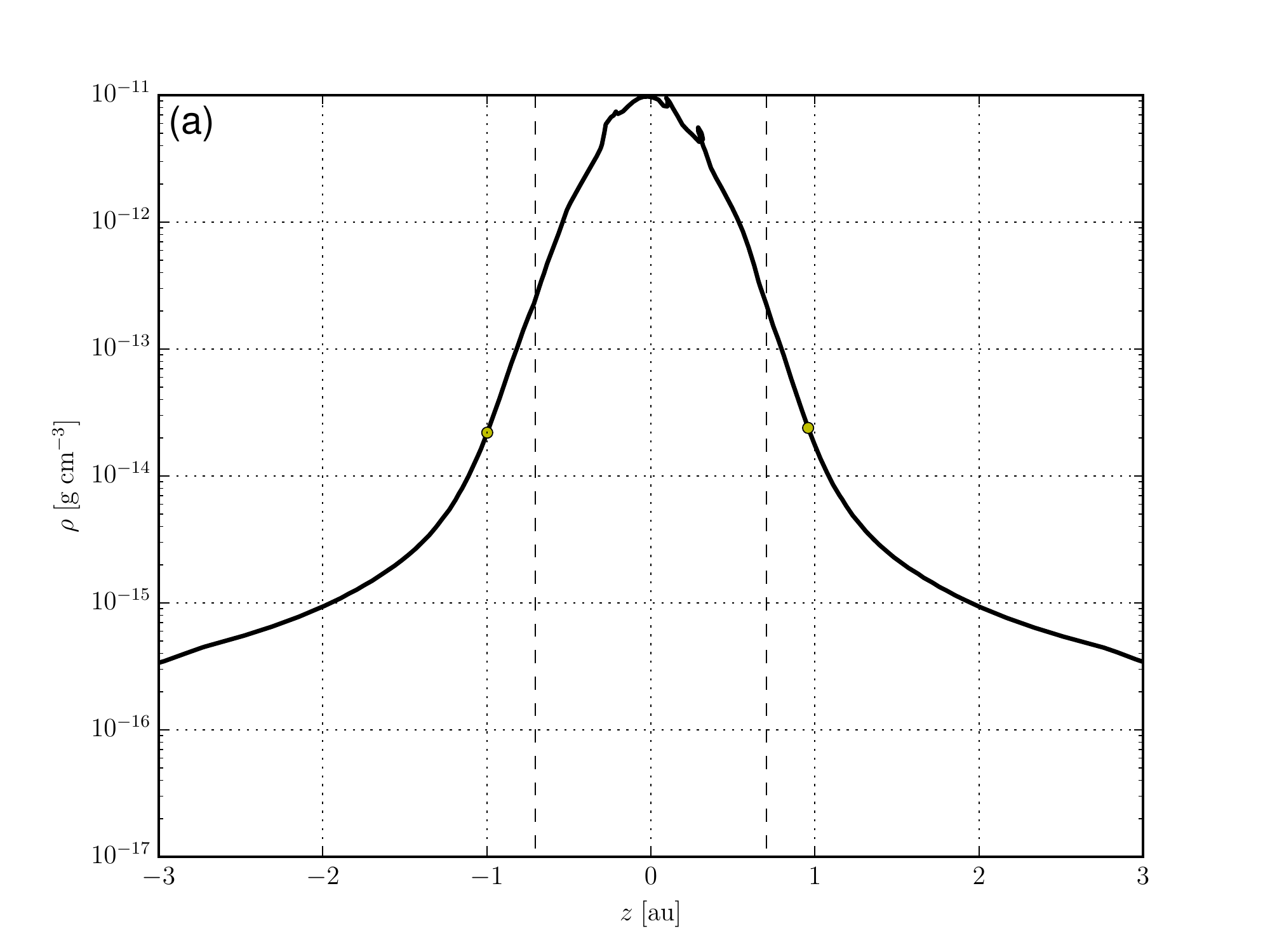}
    \includegraphics[width=1.0\columnwidth]{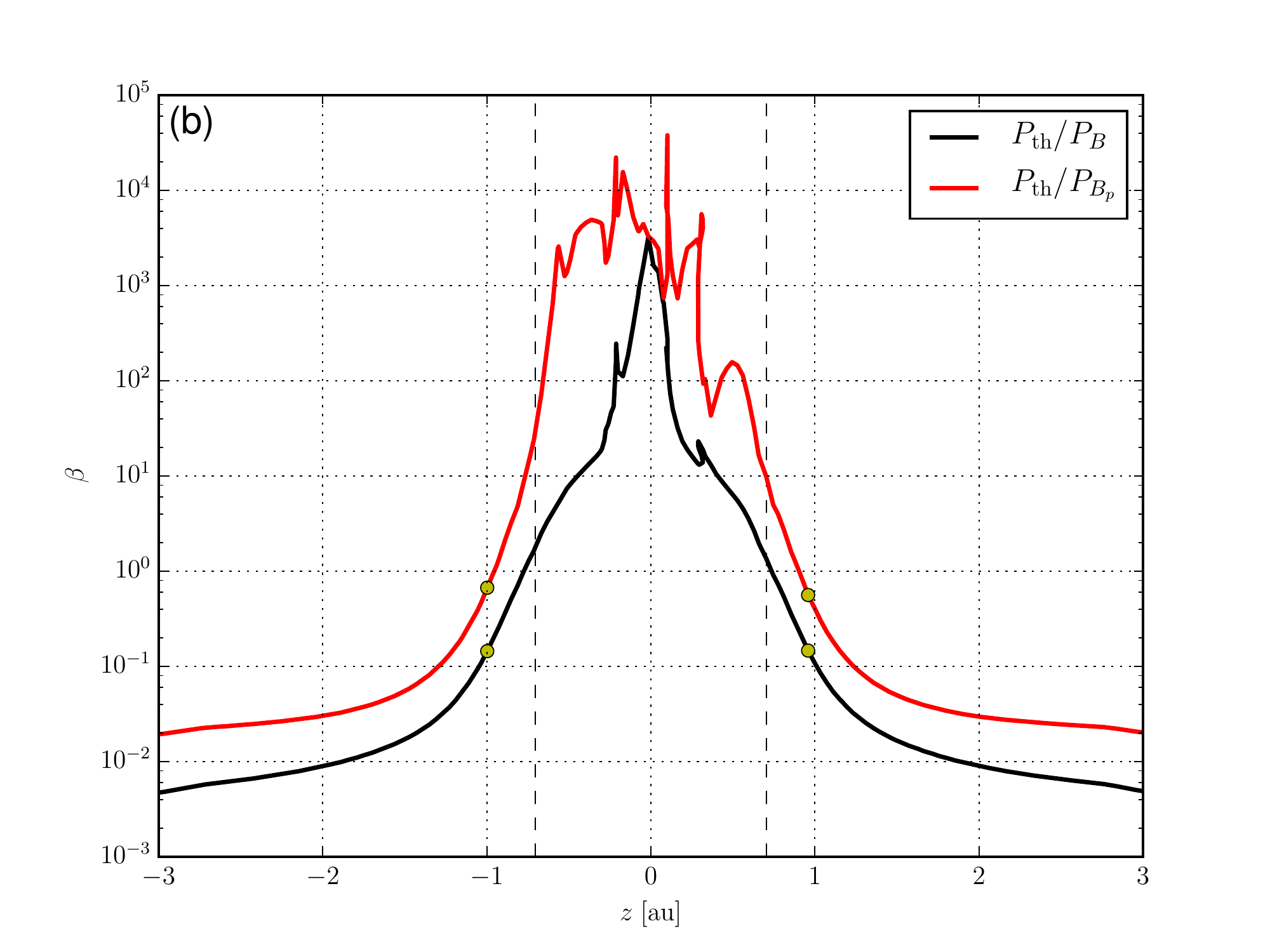}
    \includegraphics[width=1.0\columnwidth]{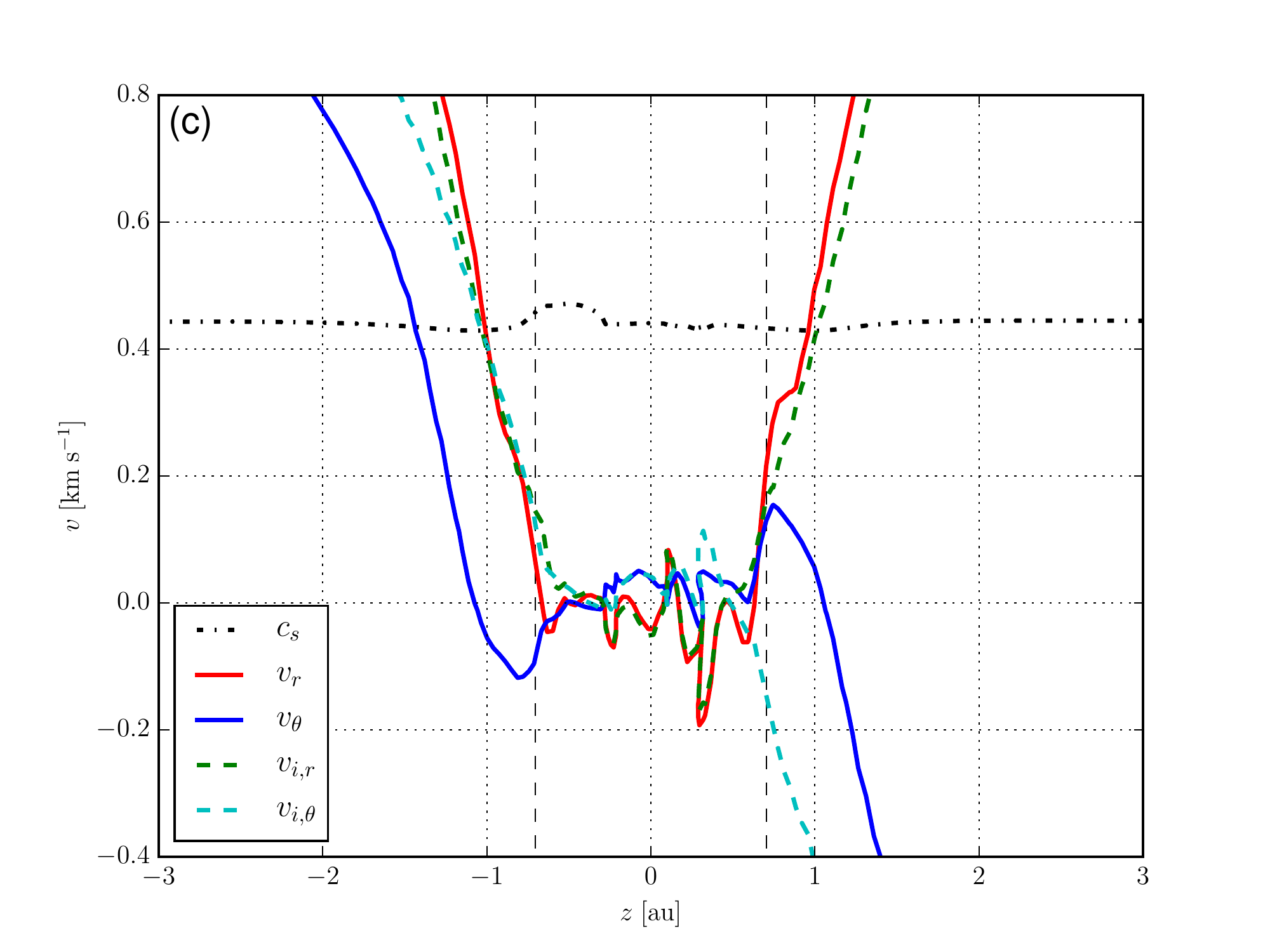}
    \includegraphics[width=1.0\columnwidth]{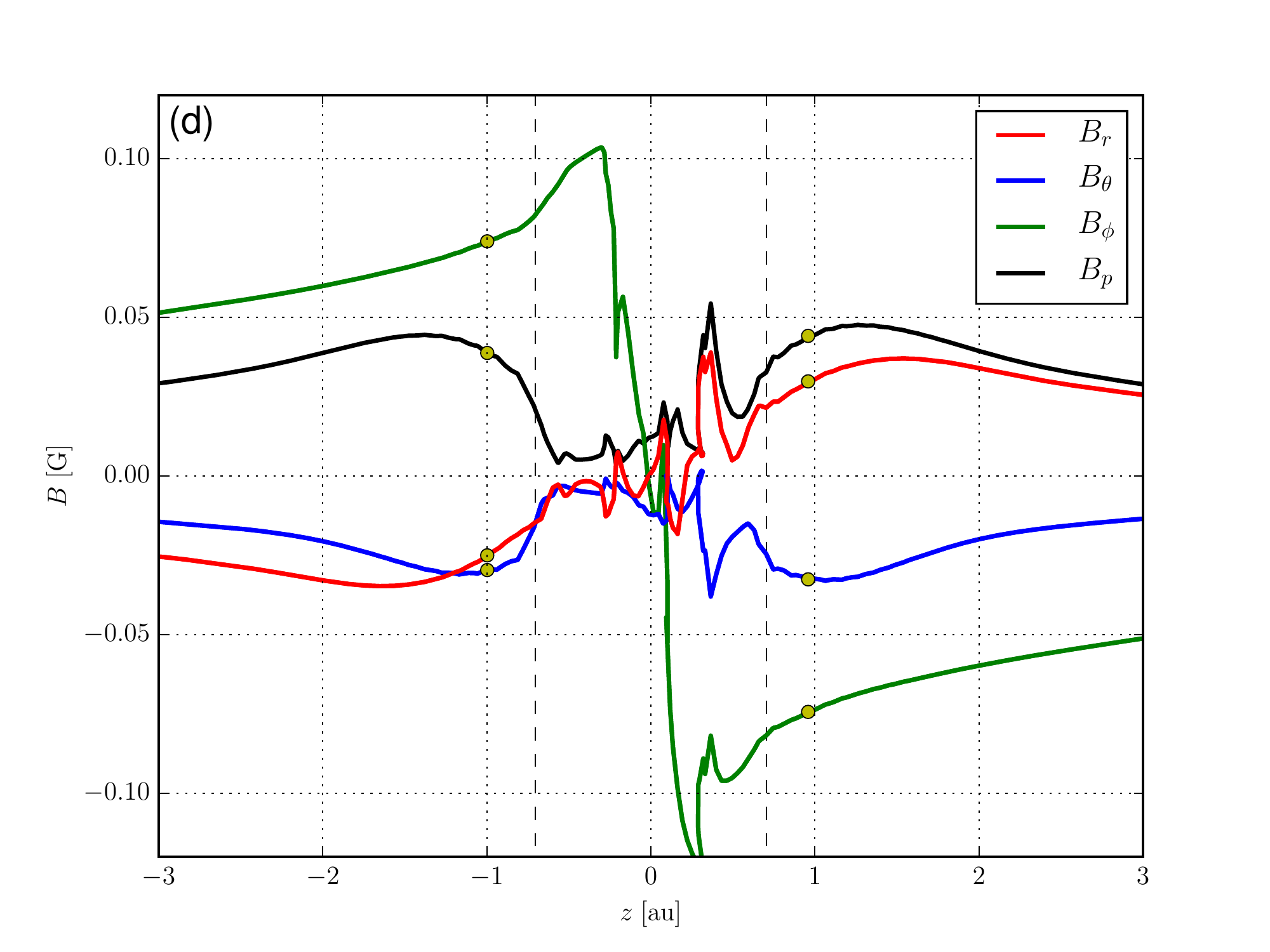}
    \caption{Physical quantities plotted along a poloidal magnetic field line as a function of the vertical height $z$. This representative field line passes through a dense ring at $r=7$~au. The panels show (a) the density distribution, (b) plasma-$\beta$ for the total magnetic field strength (black) and for the poloidal magnetic field strength (red), (c) the poloidal components of the neutral (solid lines) and ion velocities (dashed lines), and the adiabatic sound speed (dash-dotted line), and (d) the magnetic field components. The yellow circles show the sonic point (where the poloidal velocity is equal to the adiabatic sound speed) and the vertical dashed lines show the initial disk height of $z=\pm2h_0$.}
    \label{fig:path_7au}
\end{figure*}

To understand the connection between the disk region and wind region more quantitatively, we plot various physical quantities along two representative field lines. In Fig.~\ref{fig:path_8au}, we show plots along a magnetic field line that passes through a gap at a radius of 8~au on the midplane (see the magenta magnetic field line plotted in Fig.~\ref{fig:beta}d). Panel (a) of Fig.~\ref{fig:path_8au} shows that the density peaks near the midplane, at a value of approximately $10^{-12}$~\gpercmcu. It drops off quick inside the disk (the vertical dashed lines mark the initial disk surface at two scale heights). Beyond the sonic point (marked by a yellow circle on the curve), the decrease in density becomes less steep, transitioning from an exponential drop-off to a power-law decline, as expected as the approximately hydrostatic disk transitions to a supersonic wind. Even in this low density gap region, the disk is still dominated by the thermal pressure, with a plasma-$\beta$ of order 10 near the midplane (panel b). The plasma-$\beta$ decreases rapidly away from the midplane, reaching a value of order 0.1 at the sonic point; beyond the sonic point, the supersonic wind is completely magnetically dominated. This magnetic field, specifically its pressure gradient, is responsible for accelerating the outermost layer of the disk through the sonic point to produce the wind. The wind acceleration along this particular field line is illustrated in panel (c), which shows clearly the transition from slow inward motion near the disk midplane ($v_r < 0$, i.e., accretion) to fast outflow through the sonic point at approximately three disk scale heights. The sound speed remains nearly constant along the field line, which is indicative of a `cold' wind with a temperature comparable to that of the disk. Note that the outflow acceleration beyond the sonic point is rather gradual, with the velocity increasing over many disk scale heights. This is consistent with magnetically driven winds with heavy mass loading \citep{2005ApJ...630..945A}. 

In panel (c), we have plotted ion speeds together with the speeds for the bulk neutral material. The largest difference is between the velocity component $v_\theta$ and $v_{i,\theta}$, especially in the wind zone. In particular, $v_{i,\theta}$ is more positive than $v_\theta$ below the disk (negative $z$), indicating that the ions are moving faster than the neutrals {\it away from} the disk. There must be a magnetic force pointing away from the disk which drives the ion-neutral drift through ambipolar diffusion; it is the same force that accelerates the wind in the first place. The situation is similar above the disk as the magnetic force that drives the wind also moves the ions away from the disk faster than the neutrals in the negative $\theta$ direction. This force comes mostly from the toroidal component of the magnetic field, which dominates the poloidal component in the wind zone, as shown in panel (d). The gradual decrease of the toroidal component, evident in panel (d), yields a magnetic pressure gradient along the poloidal field line that lifts up the material near the disk surface against gravity and slowly accelerates it to produce a wind. Closer to the midplane, the poloidal field component (particularly $B_\theta$) becomes more dominant, indicating that the field line passes through the gap region of the disk with relatively little bending in the $r$ direction or twisting in the $\phi$ direction.

The situation is quite different along the field line that passes through a dense ring at a radius of 7~au (Fig.~\ref{fig:path_7au}). Here, the density at the midplane is an order of magnitude higher than that of the neighbouring gap (see Fig.~\ref{fig:path_8au}a). It decreases rapidly away from the midplane and the decrease becomes slower beyond the sonic point, signalling the transition from the disk to a wind. The plasma-$\beta$ through the ring is more than $10^3$ near the disk midplane, more than two orders of magnitude larger than that in the gap. In other words, as measured by $\beta$, the ring is much less magnetized than the gap. The difference is even larger when only the poloidal field component is considered. From the red curve in Fig.~\ref{fig:path_7au}(b), which plots the ratio of the thermal pressure to the magnetic pressure due to the poloidal field only, it is clear that the poloidal field in the ring is not only weak (and much weaker than the toroidal component) but also highly variable as a function of $z$. Nevertheless, the basic disk-wind structure is preserved, as shown in panel (c), where the poloidal components of the ion and neutral velocities are plotted. Again, the transition from a slowly moving disk (in the poloidal plane) to a faster expanding supersonic wind is evident. Compared to the relatively smooth accretion in the gap, which has a single negative peak in $v_r(z)$ (see red line in Fig.~\ref{fig:path_8au}c), the radial flow in the ring is much more complex: it has six negative peaks and at least three positive peaks, indicating the coexistence of multiple channels of accretion and expansion in the ring. These channels are reflected in the magnetic structure (panel d), particularly in the vertical distribution of the radial component, $B_r(z)$, which has several sign reversals consistent with channel flows in the weakly magnetized ring. As alluded to earlier, the most striking difference in the magnetic field between the ring and the gap is the strength of the poloidal magnetic field, especially the polar component, $B_\theta$, which is much lower in the ring than in the gap (compare Fig.~\ref{fig:path_8au}d and \ref{fig:path_7au}d). In the ring, the magnetic field is completely dominated by the toroidal component, except near the midplane where $B_\phi$ changes direction.

\subsection{Formation of rings and gaps}\label{sec:ring}
In this subsection, we will demonstrate that the formation of rings and gaps in the reference simulation is closely related to the magnetic structure that develops in the disk, particularly the sharp pinching of the poloidal field line near the midplane that leads to magnetic reconnection. This field pinching is caused by the development of a current sheet near the midplane, as we show next.   

\begin{figure*}
    \centering
    \includegraphics[width=1.0\columnwidth]{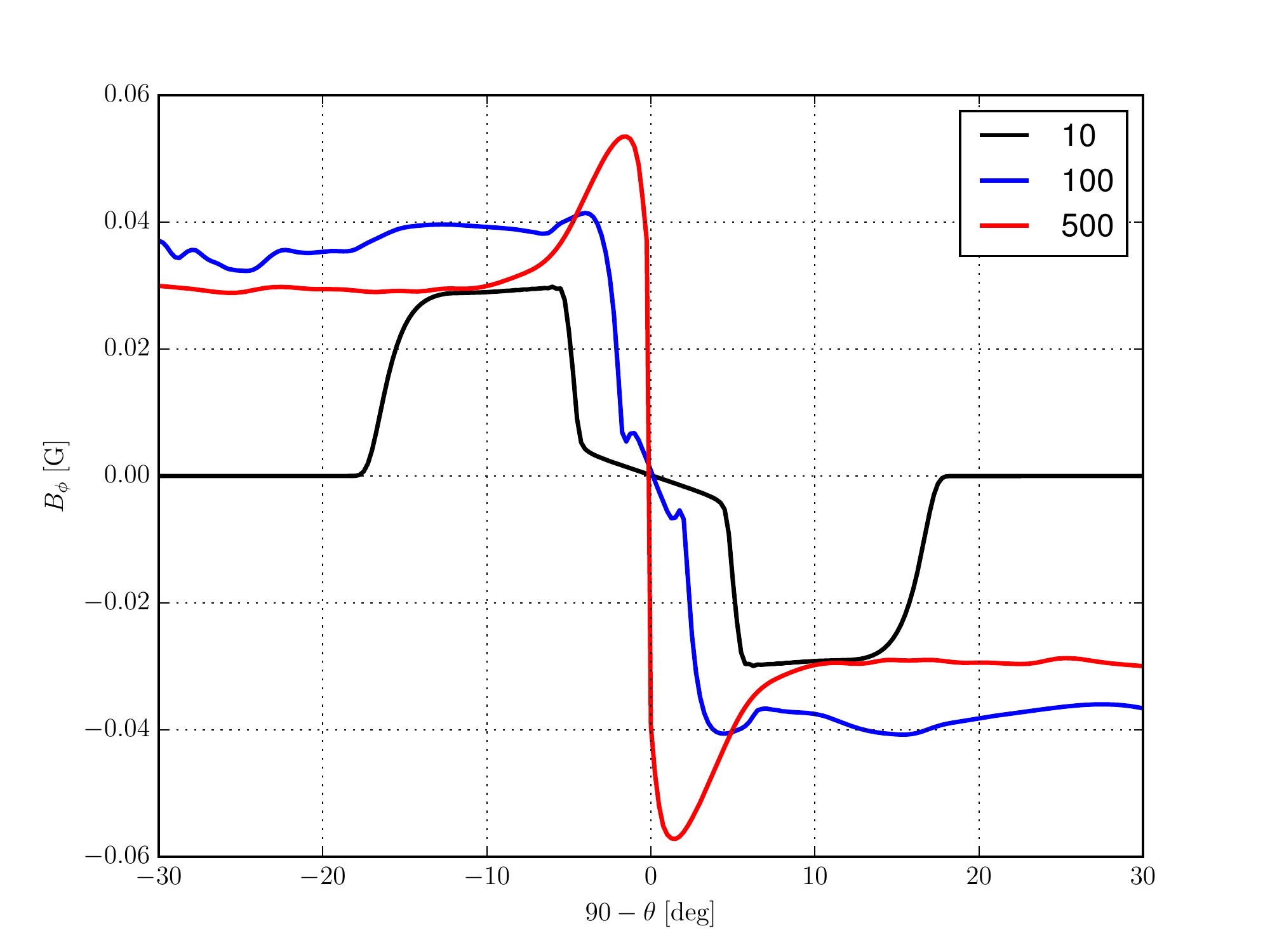}
	\includegraphics[width=1.0\columnwidth]{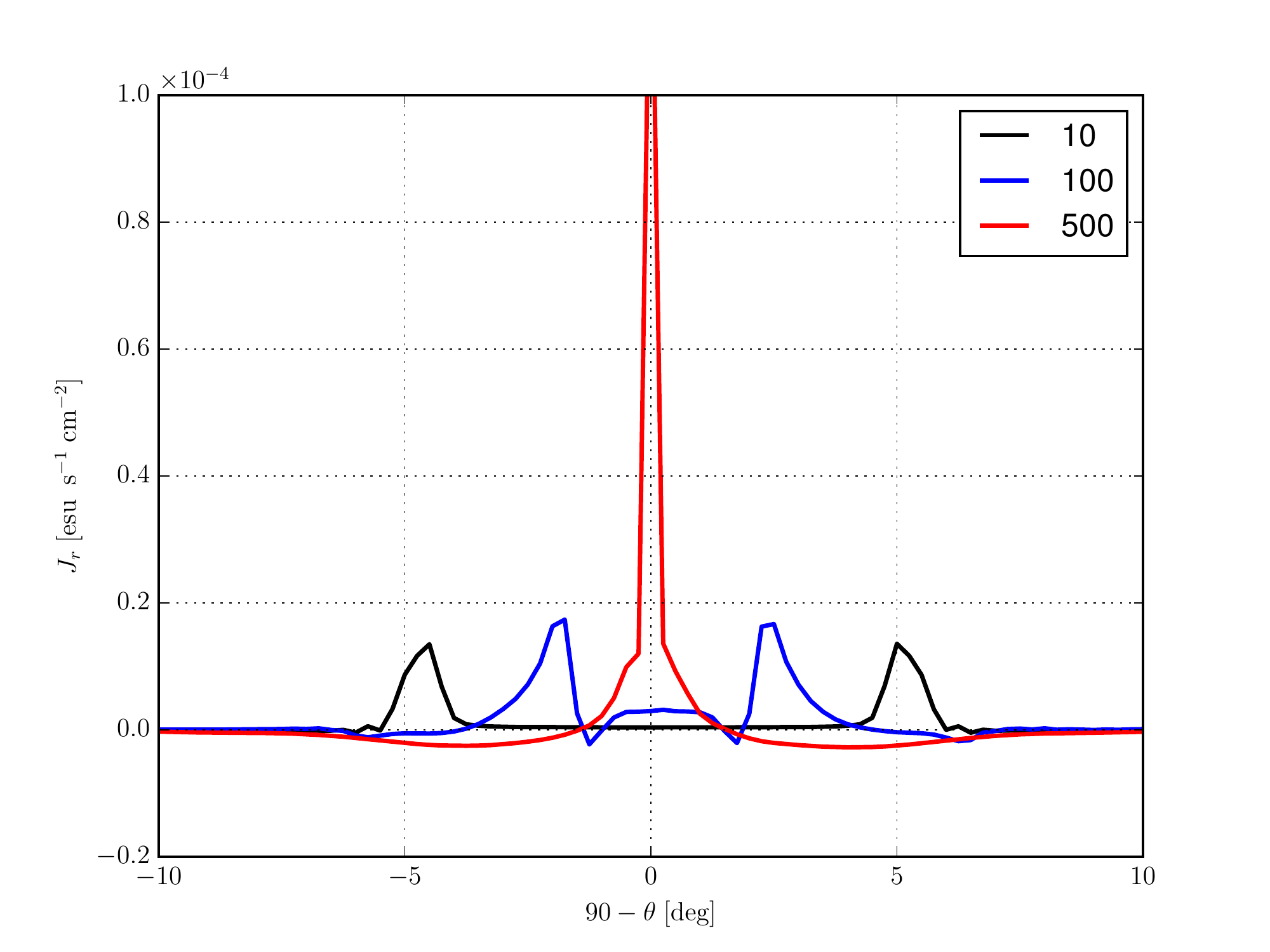}
    \caption{The toroidal magnetic field $B_\phi$ (left) and the radial current density $J_r$ (right) plotted versus $90^\circ-\theta$ (zero at the midplane and negative below it) at radius $r=20$~au. The legend labels are the simulation time in units of the inner orbital period $t_0$.}
    \label{fig:jmid}
\end{figure*}

\subsubsection{Midplane current sheet}\label{sec:jmid}
We start by reminding the reader of the simulation setup, where the disk is rotating slightly sub-Keplerian for $\vert\pi/2-\theta\vert<2\epsilon$, the coronal regions above and below the disk are not rotating, and the magnetic field has no $\phi$ component initially. As the simulation begins to run, a toroidal magnetic field is quickly generated near the boundary between the rotating disk and the stationary corona due to differential rotation. This can been seen in the left panel of Fig.~\ref{fig:jmid}, which plots the toroidal component of the magnetic field as a function of $\theta$ at a representative disk radius $r=20$~au. At $t/t_0=10$, the solid black line shows that a toroidal magnetic field has already been generated near the disk surface, but has yet to penetrate into the bulk of the disk. Associated with the variation of $B_\phi$ with polar angle $\theta$ is a radial current,
\begin{equation}
J_r = \frac{c}{4\pi}~\frac{1}{r\sin\theta}\frac{\partial(B_\phi\sin\theta)}{\partial\theta}\approx\frac{c}{4\pi}\frac{dB_\phi}{r d\theta},
\end{equation}
where $\sin\theta\approx 1$ in the thin disk. This current is plotted in the right panel of Fig.~\ref{fig:jmid}, which shows two positive peaks near $5^\circ$ above and below the disk midplane at time $t/t_0=10$, corresponding to the sharp drop of $\vert B_\phi\vert$ going from the corona into the disk. At later times, the region of high toroidal field, $\vert B_\phi\vert$, above and below the midplane expands both outward into the corona and, more importantly, toward the disk midplane. The latter drives the two current layers (one on each side of the midplane) towards the midplane, until they merge together into a single, thin, current sheet, as shown in Fig.~\ref{fig:jmid} (right). 

Ambipolar diffusion plays a key role producing the thin midplane current sheet. First, were it not for the presence of AD (or some other magnetic diffusivity, such as resistivity), prominent avalanche streams would have developed near the disk surface, which would drive the entire disk and its envelope into an unsteady state and make the formation of a laminar midplane current sheet impossible (see \citetalias{2017MNRAS.468.3850S} and Section~\ref{sec:els} below)\footnote{In this reference run, AD does not appear to suppress the development of the MRI completely. Channel-like features are evident at large radii where the disk is better coupled to the magnetic field as measured by the Elsasser number (see Fig.~\ref{fig:global}c).}. Second, as first stressed by \citet{1994ApJ...427L..91B}, AD tends to steepen the magnetic gradient near a magnetic null, i.e., where the magnetic field changes polarity. The reason is that the Lorentz force associated with the magnetic pressure gradient drives the ions (relative to the neutrals) toward the null from both sides. Since the field lines (of opposite polarity across the null) are tied to the ions, they are dragged towards the null as well, leading to a steepening of the magnetic gradient, which in turn yields a stronger magnetic force that drives the ions and the field lines even closer to the null. Since the ambipolar magnetic diffusivity, $\eta_{A}$, is proportional to the field strength (see equation~\ref{ADdiffusion}) and thus vanishes at the null, this steepening would result in an infinitely thin, singular, current sheet in principle. In practice, the thickness of the current sheet is limited by the grid resolution. 

Sharp magnetic field reversals that give rise to thin current sheets are prone to reconnection. However, this is not the case for the midplane current sheet shown in Fig.~\ref{fig:jmid}, because it is produced by the reversal of the toroidal field component $B_\phi$ and it is impossible to reconnect oppositely directing toroidal fields under the adopted axisymmetry\footnote{Reconnection of the highly pinched toroidal field is expected in 3D, and will be explored in a future investigation.}. Nevertheless, this primary current sheet leads to another current component that does allow for reconnection.  

\subsubsection{Reconnection of pinched radial magnetic field}

The secondary component of the midplane current sheet develops as a result of mass accretion in the disk, which is concentrated in the primary radial current ($J_r$) sheet near the midplane (see Fig.~\ref{fig:jmid}). The mass accretion is driven by the removal of angular momentum due to the Lorentz force ($\propto J_r B_\theta$) in the azimuthal direction, which peaks in the radial current sheet where the toroidal magnetic field changes sign. Pictorially, as the $\phi$ component of the magnetic field changes from the $+\phi$ direction below the disk to the $-\phi$ direction above the disk in a thin midplane layer, the magnetic field lines become severely kinked in the azimuthal direction. The sharp field kink generates a magnetic tension force in the $-\phi$ direction that exerts a braking torque and drives the disk accretion in the current sheet.

\begin{figure*}
    \centering
	\includegraphics[width=2.0\columnwidth]{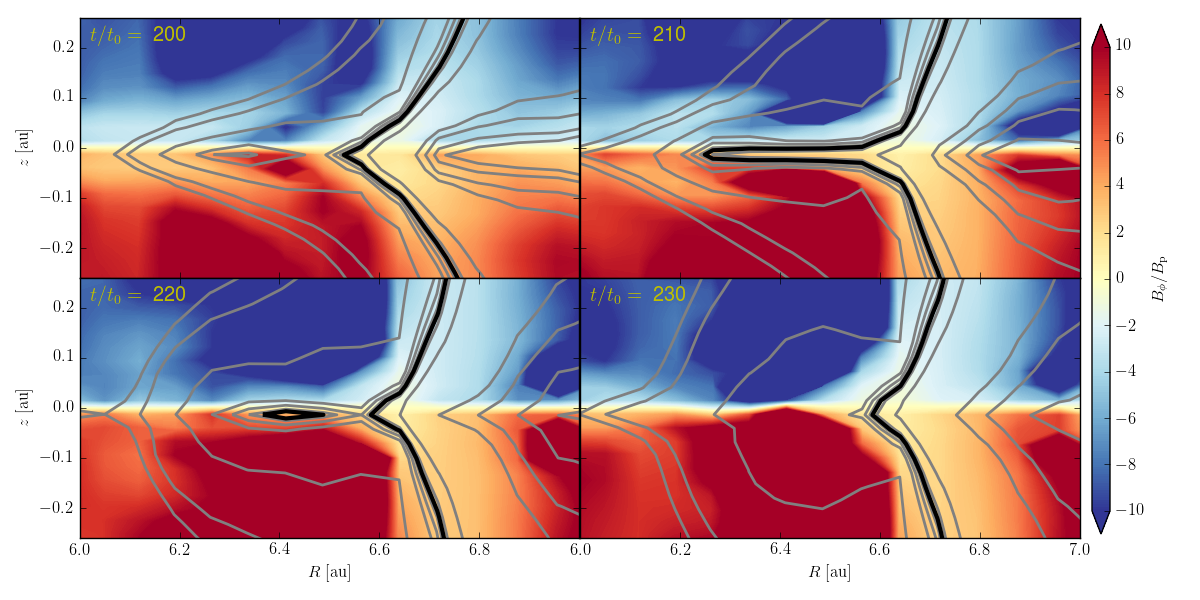}
    \caption{Poloidal magnetic field lines at four different times are shown in grey, with a reconnecting field line highlighted in black. The colours show the ratio of the toroidal to poloidal magnetic field, $B_\phi/B_\mathrm{p}$.}
    \label{fig:reconnect}
\end{figure*}

The accretion through the midplane current sheet now drags the poloidal magnetic field lines inward with the flow. This results in a pronounced radial pinch of the poloidal magnetic field, which transports the poloidal magnetic flux inward and yields another current component in the azimuthal direction, $J_\phi$. Eventually, the radial pinch becomes so severe that the magnetic field reconnects, forming poloidal magnetic field loops that are reminiscent of the tearing mode instability (\citealt{1963PhFl....6..459F}; see also \citealt{2009ARA&A..47..291Z}). An example of the reconnection process is shown in Fig.~\ref{fig:reconnect}, which plots lines of constant poloidal magnetic flux, i.e. poloidal magnetic field lines, along with the ratio of the toroidal to poloidal magnetic field, $B_\phi/B_\mathrm{p}$ (colour contours), at a radius centred on $r=6.5$~au from time $t/t_0=200$ to 230. At the first frame shown, the poloidal magnetic field highlighted in bold already has a kink across the midplane. The pinch grows with time until the field line is stretched almost parallel to the midplane over approximately four radial grid cells at $t/t_0=210$. By the next frame shown at $t/t_0=220$, the field has reconnected, forming a loop near $r=6.4$~au. In the last frame the loop has disappeared, however, the process of its formation has left a lasting mark on the magnetic field structure. The region now has a much larger toroidal magnetic field component compared to its poloidal component. The now weaker poloidal field strength can be seen clearly in the last panel from the lack of field lines in the post-reconnection region.

\subsubsection{Reconnection-driven ring and gap formation}\label{sec:formation}
The general picture for the reconnection-driven ring and gap formation is as follows. A poloidal magnetic field line that initially threads the disk rather smoothly is dragged by preferential accretion near the midplane into a highly pinched radial configuration (see upper-right panel of Fig.~\ref{fig:reconnect} for an illustration). Reconnection of the highly pinched field line produces a poloidal magnetic loop next to a poloidal field line that still threads the disk more smoothly (see lower-left panel of Fig.~\ref{fig:reconnect}). After reconnection, the material trapped on the magnetic loop is detached from the original (pre-reconnection) poloidal field line, which results in the separation of matter and (poloidal) magnetic flux. Specifically, there is no net poloidal magnetic flux passing through the matter on the loop and there is less matter remaining on the original poloidal field line (since some of the matter on the original field line is now on the detached magnetic loop). The net effect is a redistribution of the poloidal magnetic flux away from the loop-forming region into an adjacent region where the poloidal flux accumulates. Since mass accretion tends to be faster in regions with stronger poloidal magnetic fields, it is not surprising that the reconnection-induced variation of the poloidal field strength with radius would lead to a spatial variation in the mass accretion rate that would in turn lead to spatial variation in the mass (surface) density, i.e., the formation of rings and gaps. \footnote{The stronger poloidal field in a gap can also lead to a faster depletion of the local disk material via a stronger magnetically driven wind (e.g., \citealt{2000A&A...353.1115C}). However, this effect is less important compared to the faster accretion in the gap in general.}

\begin{figure}
    \centering
	\includegraphics[width=1.0\columnwidth]{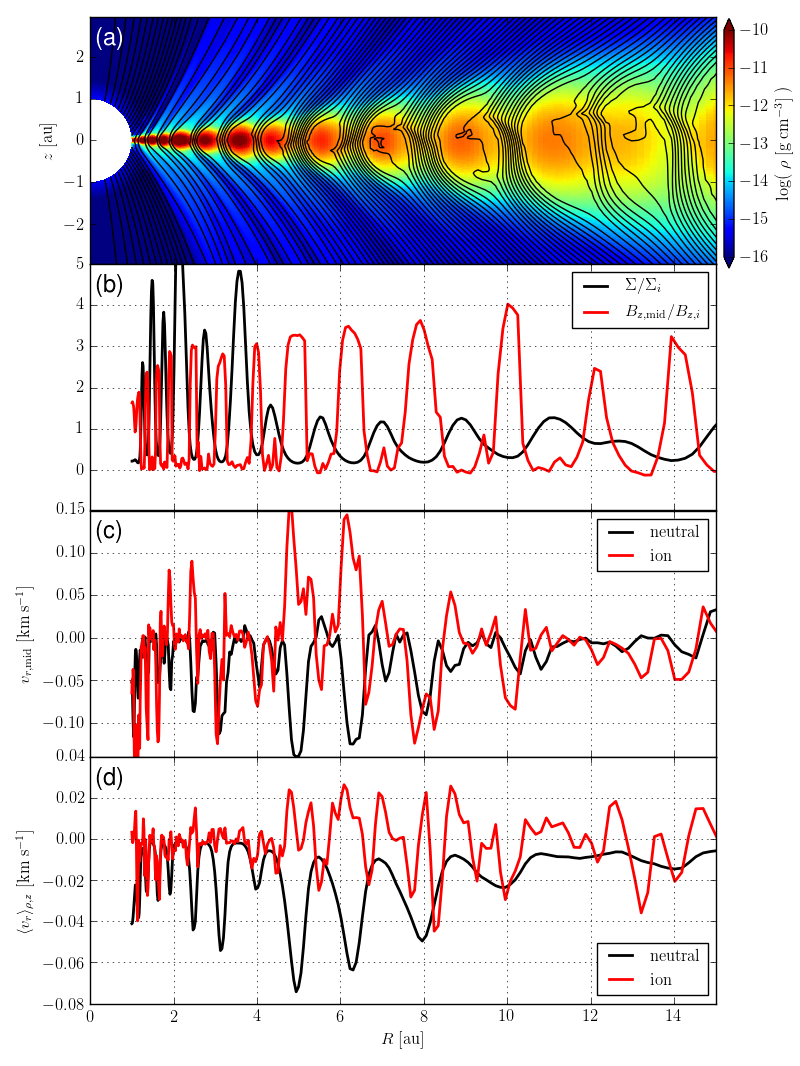}
    \caption{The density, magnetic field strength, and velocities plotted up to $r=15$~au at $t/t_0=2500$. The panels show (a) the logarithm of the density (\gpercmcu) in colour and the poloidal magnetic field lines (i.e., magnetic flux contours) in black, (b) the surface density normalized to the initial radial distribution, $\Sigma_i\propto r^{-1/2}$, and the vertical magnetic field strength at the midplane normalized to its initial distribution, $B_{z,i}\propto r^{-5/4}$, (c) the radial velocity (\kms) of neutrals (black) and ions (red) at the midplane, and (d) the density-weighted vertical average of the radial velocity (\kms) of neutrals (black) and ions (red) over $z=\pm2h$. (See the supplementary material in the online journal for an animated version of this figure.)}
    \label{fig:bunch}
\end{figure}

\begin{figure}
    \centering
	\includegraphics[width=1.0\columnwidth]{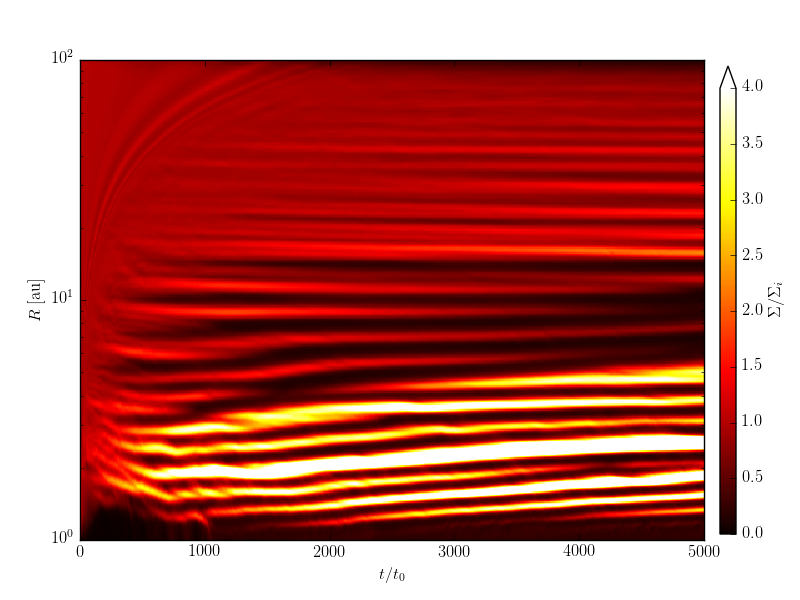}
    \caption{The surface density of the disk (normalized to its initial radial distribution) as a function of radius and time, showing that most of rings and gaps created in the reference run remain stable for thousands of inner orbital periods.}
    \label{fig:surfden_v_time}
\end{figure}

The preferential concentration of poloidal field lines inside the low-density gaps of the disk is illustrated pictorially in Fig.~\ref{fig:bunch}(a), which is a snapshot of the inner part of the reference simulation (up to 15~au) at a representative time when the simulation has reached a quasi-steady state, $t/t_0=2500$. The field concentration is quantified in panel (b), which plots the vertical component of the magnetic field ($B_z$, in red) at the midplane and surface density of the disk (in black), both relative to their initial values at $t=0$. Two features stand out: (1) there is a strong anti-correlation between the surface density and vertical field strength, as expected in the reconnection-induced scenario of ring and gap formation, and (2) the poloidal magnetic field is typically much weaker in the high-density ring regions (where $B_z$ is close to zero) than in the low-density gap regions (where $B_z$ is increased over its initial value by a factor of $2-4$). This drastic segregation of poloidal magnetic flux relative to matter appears to be quite stable with time (persisting for thousands of inner orbits), despite the fact that there is continued mass accretion through both the rings and gaps.

For our reference run, we believe that the key to maintaining the concentration of poloidal magnetic flux in the gap regions is ambipolar diffusion: it allows the bulk neutral material to accrete radially inward through the gaps without dragging the ions (and the magnetic field lines tied to them) with it. This is illustrated in panel (c) of Fig.~\ref{fig:bunch}, which plots the radial component of the ion (red line) and neutral (black line) velocity at the midplane. As expected, the neutrals accrete faster in the low-density gaps than in the high-density rings. One may naively expect the same trend for the ions but, in the presence of significant ambipolar diffusion, this is not necessarily the case. Indeed, at the time shown in Fig.~\ref{fig:bunch}, the ions are moving {\it outward} in several of the gaps, especially the two near $r=5$ and 6.5~au. The ions are forced to expand against the infalling neutrals by an outward Lorentz force due to either a temporary poloidal flux concentration near the inner edge of the gap (in a manner that is reminiscent of the forced ion-neutral separation in the AD-shock in magnetized accretion onto low-mass protostars first described in \citealt{1996ApJ...464..373L}) or an outward magnetic pressure gradient from the toroidal field component. In any case, the radially outward Lorentz force in the gap appears strong enough to keep the ions (and the poloidal field lines) in a state of dynamic equilibrium against the rapid infall of neutrals, at least under the assumption of (2D) axisymmetry. The dynamic equilibrium of the ions (and the field lines attached to them) is shown more clearly in Fig.~\ref{fig:bunch}(d), which plots the vertically averaged radial velocity weighted by density. The average ion speed fluctuates around zero as the neutrals accrete inward, especially in the gap regions. Whether this remains true in full 3D simulations is unclear and will be explored in future investigations.

The rings and gaps, once fully developed, remain remarkably stable over time. This is illustrated in Fig.~\ref{fig:surfden_v_time}, where the surface density (relative to its initial distribution) is plotted at each radius as a function of time, as done in \citet{2017A&A...600A..75B} for plasma-$\beta$ (see their Fig.~30). Note that most of the rings and gaps are stable for at least 4000 inner orbital periods. There are a few exceptions. For example, the two rings near 10~au appear to merge together around $t/t_0=5000$, whereas the ring at 6~au starts to fade away at later times. It would be interesting to determine whether these rings and gaps remain stable for long periods of time in full 3D simulations.

\begin{table*}
  \centering
  \caption{Model parameters for all simulation runs.}
  \label{tab:sims}
  \begin{tabular}{l c c c c c c c c}
	\hline
    \hfill & $\beta/10^3$ & $\gamma/10^{-3}$ & $\Lambda_0=\gamma\rho_{i,0}/\Omega_0$ & $\eta_{A,0}/10^{14}$ & $\eta_O/10^{14}$ \\
    \hfill & & $[\rm{cm}^3~\rm{g}^{-1}~\rm{s}^{-1}]$ & & $[\rm{cm}^2~\rm{s}^{-1}]$ & $[\rm{cm}^2~\rm{s}^{-1}]$ \\
    \hline
    ad-els0.01      & 0.922 & 0.1763 & 0.01 & 243   & --  \\
    ad-els0.05      & 0.922 & 0.8816 & 0.05 & 48.6  & --  \\
    ad-els0.1       & 0.922 & 1.763  & 0.1  & 24.3  & --  \\
    ad-els0.25~(ref)& 0.922 & 4.408  & 0.25 & 9.71  & --  \\
    ad-els0.5       & 0.922 & 8.816  & 0.5  & 4.86  & --  \\
    ad-els1.0       & 0.922 & 17.63  & 1.0  & 2.43  & --  \\
    ad-els2.0       & 0.922 & 35.26  & 2.0  & 1.21  & --  \\
    ideal           & 0.922 & --     & --   & --    & --  \\
    oh0.26          & 0.922 & 4.408  & 0.25 & 9.71  & 2.5 \\
    oh2.6           & 0.922 & 4.408  & 0.25 & 9.71  & 25  \\
    oh26            & 0.922 & 4.408  & 0.25 & 9.71  & 250 \\
    beta1e2         & 0.0922& 4.408  & 0.25 & 97.1  & --  \\
    beta1e4         & 9.22  & 4.408  & 0.25 & 0.971 & --  \\
    \hline
  \end{tabular}
\end{table*}

\section{Effects of magnetic coupling and field strength on ring and gap formation}\label{sec:param}
The most important features in the reference simulation are the rings and gaps that develop spontaneously in the disk. In this section, we will explore how their formation is affected by how well the ions (and therefore the magnetic field) are coupled to the bulk neutral fluid. The magnetic field coupling is changed by varying the AD Elsasser number at $r_0$ and $\theta=\pi/2$, $\Lambda_0$, which sets the scale for the Elsasser number everywhere; the ion density profile is unchanged. The AD Elsasser number controls the coupling between the ions and the bulk neutral fluid, as $\Lambda\propto\gamma\rho_i$, where $\gamma\rho_i$ is the collision frequency between the neutrals and ions. When $\Lambda$ is small, so is the collision frequency, and, therefore, the ions/magnetic field are poorly coupled to the neutral fluid. As $\Lambda$ increases, the magnetic field becomes increasingly coupled to the motions of the bulk neutral fluid. The ideal MHD limit is reached as the AD Elsasser number approaches infinity, $\Lambda\rightarrow\infty$. In Table~\ref{tab:sims}, we list the simulation runs performed to examine the effect that the magnetic coupling strength has on the ring and gap formation mechanism described in the previous section. First, we will present the results of the simulations as the Elsasser number increases from 0.01 to 2 (in the simulations named ad-els0.01, ad-els0.05, ad-els0.1, ad-els0.5, ad-els1.0, and ad-els2.0), as well as the ideal MHD case (Section~\ref{sec:els}). Next, we show the effects of introducing an explicit Ohmic resistivity into the reference simulation, where the Ohmic resistivity, $\eta_O$, is constant everywhere and is equal to 0.26, 2.6, and 26 times the initial effective ambipolar resistivity, $\eta_{A,0}$, at $r_0$ on the disk midplane ($\theta=\pi/2$). These simulations are named oh0.26, oh2.6, and oh26 respectively. They are discussed in Section~\ref{sec:beta} together with simulations that have different initial magnetic field strengths, with the midplane plasma-$\beta$ approximately ten times higher (model beta1e4) and lower (beta1e2) than that of the reference run. We conclude this section with an analysis of the magnetic stresses in the disk (Section~\ref{sec:stress}).

\begin{figure*}
    \centering 
	\includegraphics[width=2.0\columnwidth]{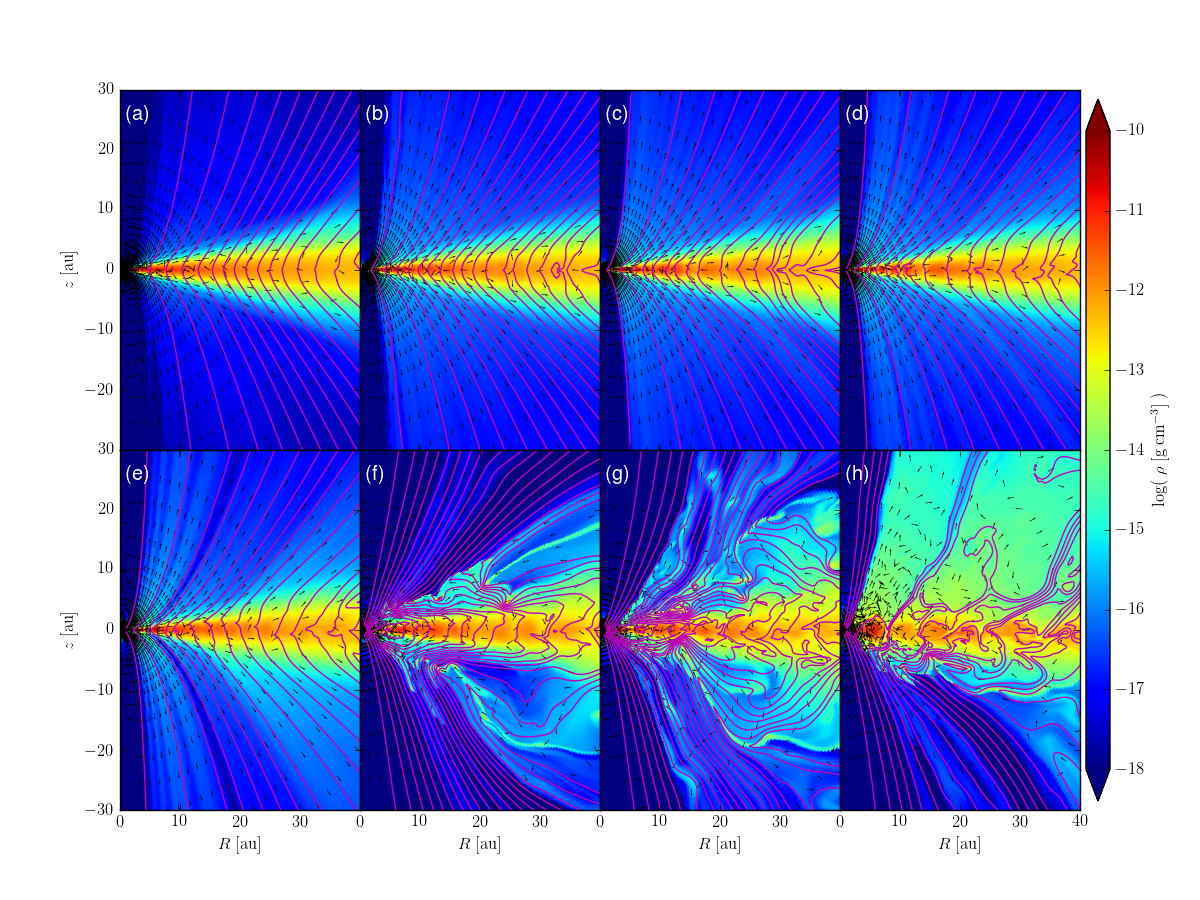}
    \caption{Snapshots at $t/t_0=2000$ of the eight simulations where the AD Elsasser number is varied. Shown is the mass volume density (logarithmically spaced colour contours in units of \gpercmcu), the poloidal magnetic field lines (magenta), and the poloidal velocity unit vectors (black). The AD Elsasser number increases sequentially from panels (a)-(h). The reference simulation (ad-els0.25) is shown in panel (d). The simulation panels in alphabetical order are: (a) ad-els0.01; (b) ad-els0.05; (c) ad-els0.1; (d) ad-els0.25; (e) ad-els0.5; (f) ad-els1.0; (g) ad-els2.0; (h) ideal. See Table~\ref{tab:sims} for details. (See the supplementary material in the online journal for an animated version of this figure.)}
    \label{fig:ad_panels}
\end{figure*}

\subsection{AD Elsasser number}\label{sec:els}
Before describing the results of the simulations, we will briefly describe our expectations as the AD Elsasser number is varied in the disk. In the reference run, we see a rather steady disk wind launched as disk material concentrates into rings and poloidal magnetic flux concentrates into gaps. The region demagnetized of poloidal field (where the density will grow to form a ring) develops as the radial magnetic field is stretched towards the $-r$ direction due to rapid accretion in the primary midplane current layer ($J_r$). As discussed in Section~\ref{sec:jmid}, the development of a strong midplane current layer where $B_\phi=0$ is a direct result of ambipolar diffusion because it is formed as the ions and toroidal magnetic field lines drift towards the magnetic null ($B_\phi=0$) relative to the bulk neutral material. In the limiting case that the Elsasser number goes to infinity, i.e., the ideal MHD case, the AD-enabled midplane current layer is not expected to develop and this ring and gap formation mechanism would be turned off. Instead, the so-called `avalanche' accretion streams are expected to develop near the disk surface, which may form rings and gaps through another mechanism (see \citetalias{2017MNRAS.468.3850S}). In the other limiting case where the Elsasser number approaches zero, the ions have no knowledge of the bulk fluid and the magnetic field will straighten out vertically without any effect on the disk at all. In particular, the field will neither launch a disk wind nor create any disk substructure. Therefore, there must be a minimum Elsasser number below which the formation of rings and gaps is expected to be suppressed.

\begin{figure*}
    \centering 
	\includegraphics[width=2.0\columnwidth]{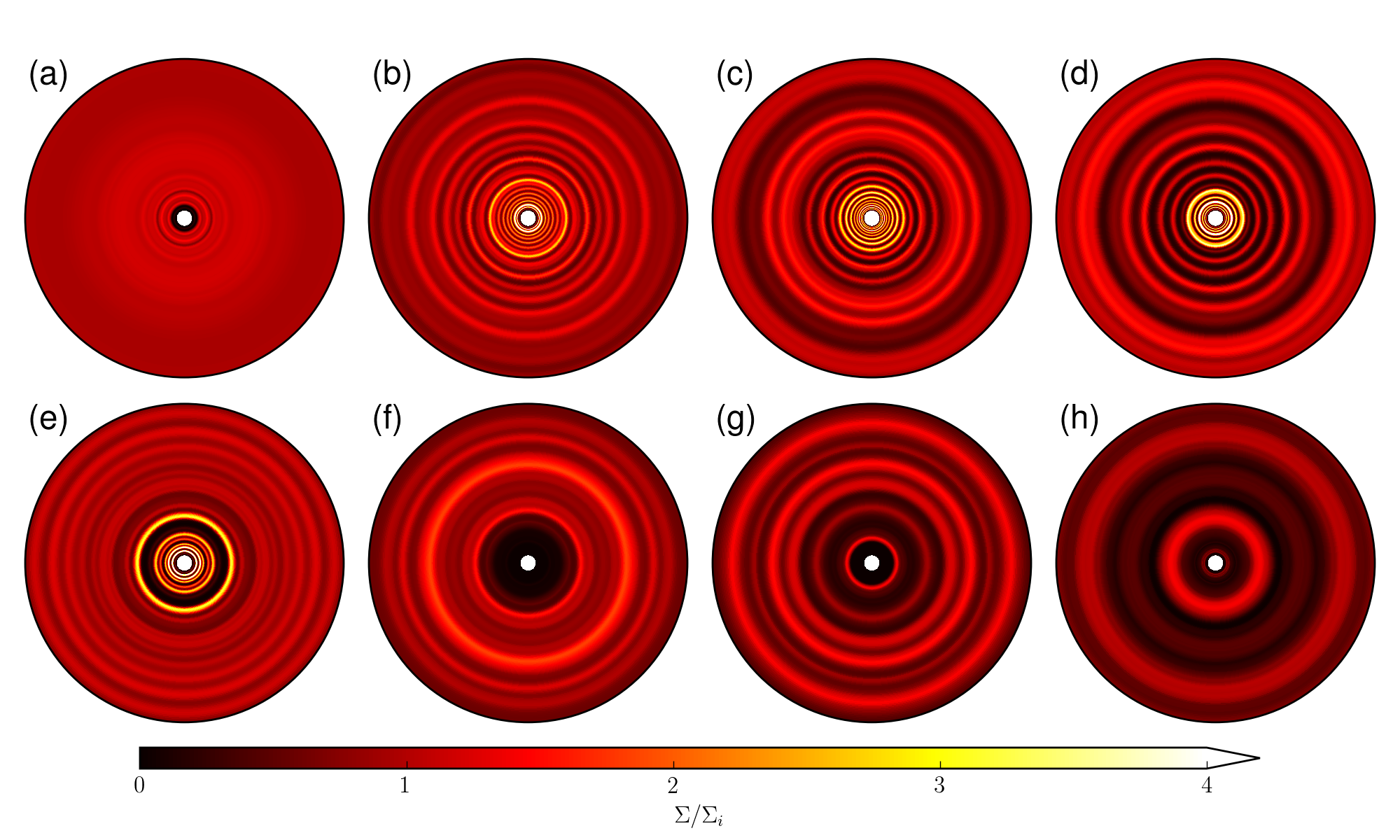}
    \caption{Face-on surface density profiles (up to a radius of 20~au) of the eight simulations where the AD Elsasser number is varied at $t/t_0=2000$. The AD Elsasser number increases sequentially from panels (a)-(h). The reference simulation (ad-els0.25) is shown in panel (d). The simulation panels in alphabetical order are: (a) ad-els0.01; (b) ad-els0.05; (c) ad-els0.1; (d) ad-els0.25; (e) ad-els0.5; (f) ad-els1.0; (g) ad-els2.0; (h) ideal. See Table~\ref{tab:sims} for details. (See the supplementary material in the online journal for an animated version of this figure.)}
    \label{fig:face-on}
\end{figure*}

These expectations are borne out by the simulations. Figures~\ref{fig:ad_panels} and \ref{fig:face-on} show, respectively, the edge-on and face-on view of simulations at a common time $t/t_0=2000$ with increasing AD Elsasser numbers from panel (a) to (h). In the most magnetically diffusive case ($\Lambda_0=0.01$), a wind is launched steadily from the disk, removes angular momentum from the disk and drives disk accretion that is rather laminar (Fig.~\ref{fig:ad_panels}a), but there is little evidence for the development of rings and gaps (Fig.~\ref{fig:face-on}a). Specifically, there is little evidence for the sharp radial pinching of poloidal magnetic field lines near the midplane that is conducive to reconnection, which is the driver of the redistribution of the poloidal magnetic flux relative to matter and is crucial to the ring and gap formation in the scenario discussed in Section~\ref{sec:ring}. The lack of sharp radial pinching is to be expected, because it would be smoothed out quickly by the large magnetic diffusivity. As the diffusivity decreases (i.e., the magnetic field becomes better coupled to the bulk disk material), it becomes easier for the midplane mass accretion to drag the poloidal field lines into a radially pinched configuration that is prone to reconnection. Indeed, this occurs over at least one decade in the AD Elsasser number, from $\Lambda_0=0.05$ to 0.5, where the disk wind remains rather steady (see Fig.~\ref{fig:ad_panels}b-e), but repeated field pinching and reconnection have created multiple rings and gaps in the disk (Fig.~\ref{fig:face-on}b-e).

\begin{figure}
    \centering
	\includegraphics[width=1.0\columnwidth]{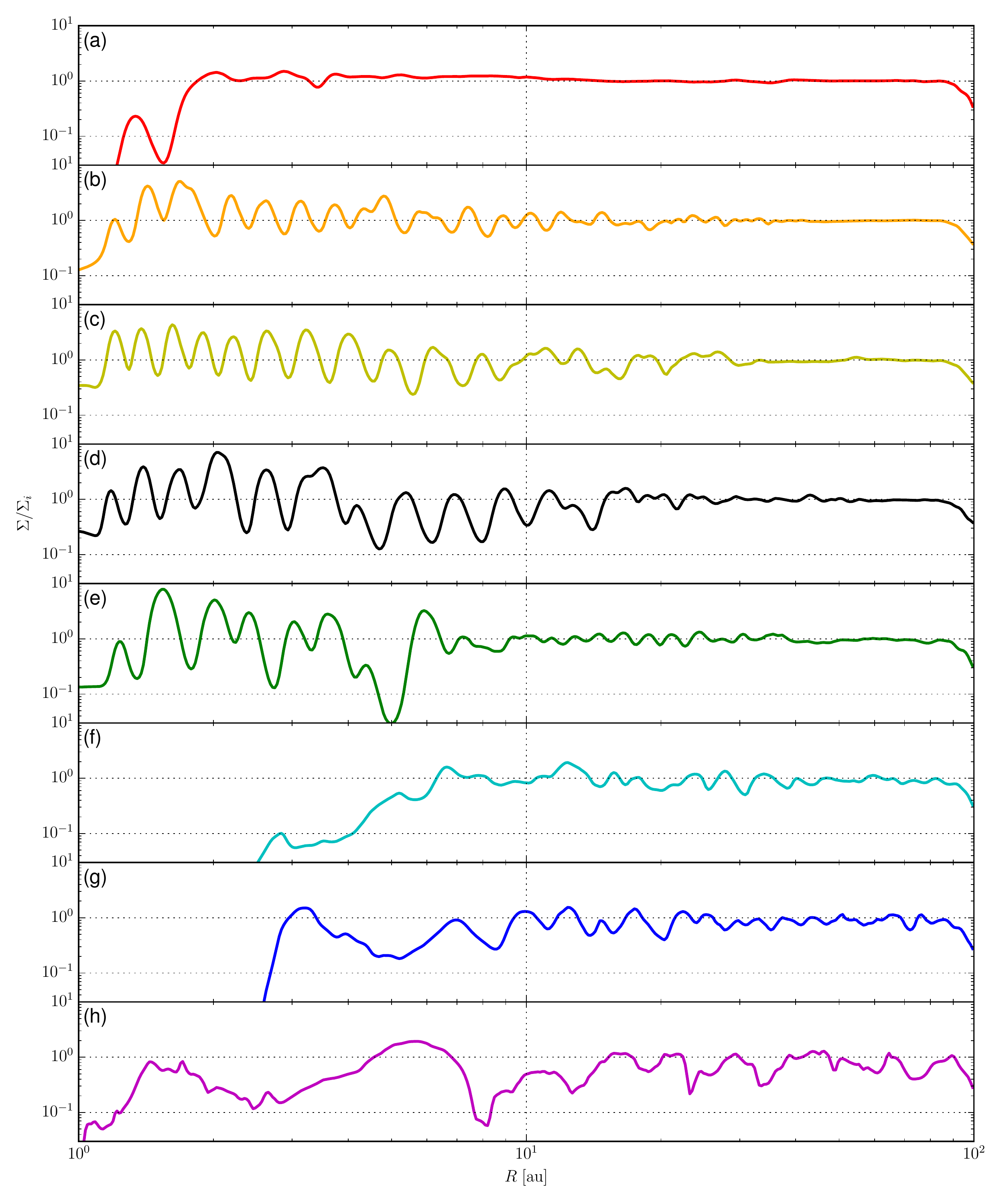}	
	\caption{Surface density profiles at time $t/t_0=2000$ for simulations with different AD Elsasser numbers. The surface density profiles are normalized to the initial surface density distribution, $\Sigma_i = \Sigma_0 (r/r_0)^{-1/2}$. The AD Elsasser number increases sequentially from the top panel to the bottom panel: (a) ad-els0.01; (b) ad-els0.05; (c) ad-els0.1; (d) ad-els0.25 (ref); (e) ad-els0.5; (f) ad-els1.0; (g) ad-els2.0; (h) ideal.}
    \label{fig:surfden1D}
\end{figure}

In the intermediate parameter regime between AD Elsasser number $\Lambda_0=0.05$ and 0.5, rings and gaps are more prominent in the inner part of the disk than in the outer part. This is quantified in Fig.~\ref{fig:surfden1D}, where the surface density of the disk (normalized to its initial value) is plotted as a function of radius. It is clear from panels (b)-(e) that most of the rings and gaps of high contrast are confined to a radius of order 10~au. One reason may be that the inner part of the disk has had more time (relative to its orbital period) for the substructures to develop. Another is that mass accretion, especially during the initial adjustment before a quasi-steady state is reached, may have redistributed some poloidal magnetic flux from the outer part of the disk to the inner part, making it easier to form rings and gaps there (the effects of magnetic field strength will be discussed in the next subsection). As the disk becomes better magnetically coupled (going from panel b to e), the number of rings and gaps in the inner 10~au region appears to decrease somewhat and the contrast between adjacent rings and gaps tends to increase on average. The higher contrast may be related to the fact that a better magnetic coupling would allow more poloidal magnetic flux to be trapped in the inner disk region. 

As the AD Elsasser number increases further (to $\Lambda_0=1.0$ and larger), another feature starts to become important. It is the development of the classical `channel flows' of the  magnetorotational instability. The channels flows are already present in the intermediate regime for $\Lambda_0$, especially in the outer part of the disk (see panels c and d of Fig.~\ref{fig:global} for the reference run), where the magnetic field is better coupled to the disk compared to the inner part as measured by the Elsasser number, which scales with radius as $\Lambda\propto r^{3/4}$. Their growth was kept in check by ambipolar diffusion in relatively diffusive models with $\Lambda_0$ up to 0.5 (panel e in Fig.~\ref{fig:ad_panels}, \ref{fig:face-on}, and \ref{fig:surfden1D}). For better magnetically coupled disks, these channel flows run away, especially near the disk surface, forming the so-called avalanche accretion streams (see \citealt{1996ApJ...461..115M,1998ApJ...508..186K} and \citetalias{2017MNRAS.468.3850S}). When fully developed, they dominate the dynamics of both the disk and the wind, driving both to an unsteady state (see panels f-h of Fig.~\ref{fig:ad_panels}).

Despite the transition to a more chaotic dynamical state, prominent rings and gaps are still formed, especially in the outer part of the disk (see panels f-h of Fig.~\ref{fig:face-on}). Part of the reason is that the strong variability of the clumpy wind is able to create variation in the disk surface density. Another, perhaps more important, reason is that the distribution of the poloidal magnetic flux is highly inhomogeneous in the disk and regions with concentrated magnetic flux tend to accrete faster leading to lower surface densities (i.e., gaps), similar to the more magnetically diffusive cases (e.g., the reference run and Fig.~\ref{fig:bunch}). Since the poloidal field bunching is present even in the ideal MHD case,\footnote{In the ideal MHD case, the wind is significantly stronger in the upper hemisphere than in the lower hemisphere (see Fig.~\ref{fig:ad_panels}h). Such an asymmetry has been observed in the non-ideal shearing box simulations of \citet{2014A&A...566A..56L} and \citet{2015ApJ...798...84B}, and in the global non-ideal MHD simulations of \citet{2017A&A...600A..75B}. The fact that it shows up in global ideal MHD simulations as well indicates that it may be a general feature of magnetically coupled disk-wind systems that should be examined more closely.} its formation does not require ambipolar diffusion, which is formally different from the more diffusive reference case\footnote{Poloidal field bunching in the ideal MHD limit has been observed in the shearing box simulations of \citet{2012A&A...548A..76M} in the case of strong disk magnetization corresponding to plasma-$\beta$ of order unity, however, artificial injection of matter onto the field lines (to prevent rapid depletion of disk material) complicates the interpretation of the result. Current global 3D ideal MHD simulations of weak field cases of $\beta\sim 10^3$ or larger (e.g., \citealt{2017arXiv170104627Z}) do not appear to show as prominent poloidal field bunching as our 2D (axisymmetric) case. Whether this difference is due to the difference in dimensionality of the simulations or some other aspects (e.g., initial and boundary conditions) remains to be determined.} (see discussion in Section~\ref{sec:formation}). Nevertheless, there is widespread reconnection in these better coupled cases as well (this is best seen in the movie version of Fig.~\ref{fig:ad_panels} available online). The reconnection is still driven by sharp radial pinching of the poloidal field lines. The difference is that here the pinching is caused by the non-linear development of unstable channel flows \citep{2014ApJ...796...31B} rather than the AD-driven midplane current sheet. As a result, the reconnection occurs more sporadically and is less confined to the midplane. The net result is the same: a redistribution of poloidal magnetic flux relative to the matter, creating regions of stronger (poloidal) magnetization that tend to form gaps and regions of weaker (poloidal) magnetization that tend to form rings. These considerations strengthen the case for reconnection as a key to ring and gap formation in a coupled, magnetized disk-wind system, either through an AD-driven midplane current sheet in relatively diffusive disks, the non-linear development of MRI channel flows in better coupled disks, or some other means. 

\begin{figure*}
    \centering 
	\includegraphics[width=2.0\columnwidth]{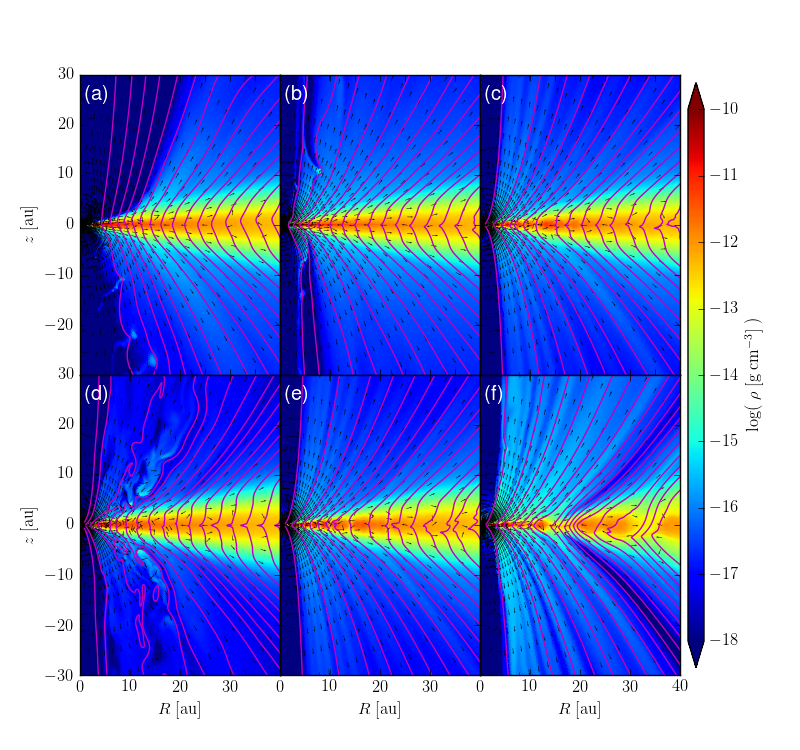}
    \caption{Snapshots of simulations where the explicit Ohmic resistivity and plasma-$\beta$ are varied at $t/t_0=2000$. Shown is the mass volume density (logarithmically spaced colour contours in units of \gpercmcu), the poloidal magnetic field lines (magenta), and the poloidal velocity unit vectors (black). In the top row, the explicit resistivity is decreased from panels (a)-(c). Plasma-$\beta$ varies from high to low across the bottom row in panels (d)-(f). The reference simulation (ad-els0.25) is shown in panel (e). The simulation panels in alphabetical order are: (a) oh26; (b) oh2.6; (c) oh0.26; (d) beta1e4; (e) ad-els0.25; (f) beta1e2. See Table~\ref{tab:sims} for details. (See the supplementary material in the online journal for an animated version of this figure.)}
    \label{fig:beta_panels}
\end{figure*}

\begin{figure*}
    \centering 
	\includegraphics[width=2.0\columnwidth]{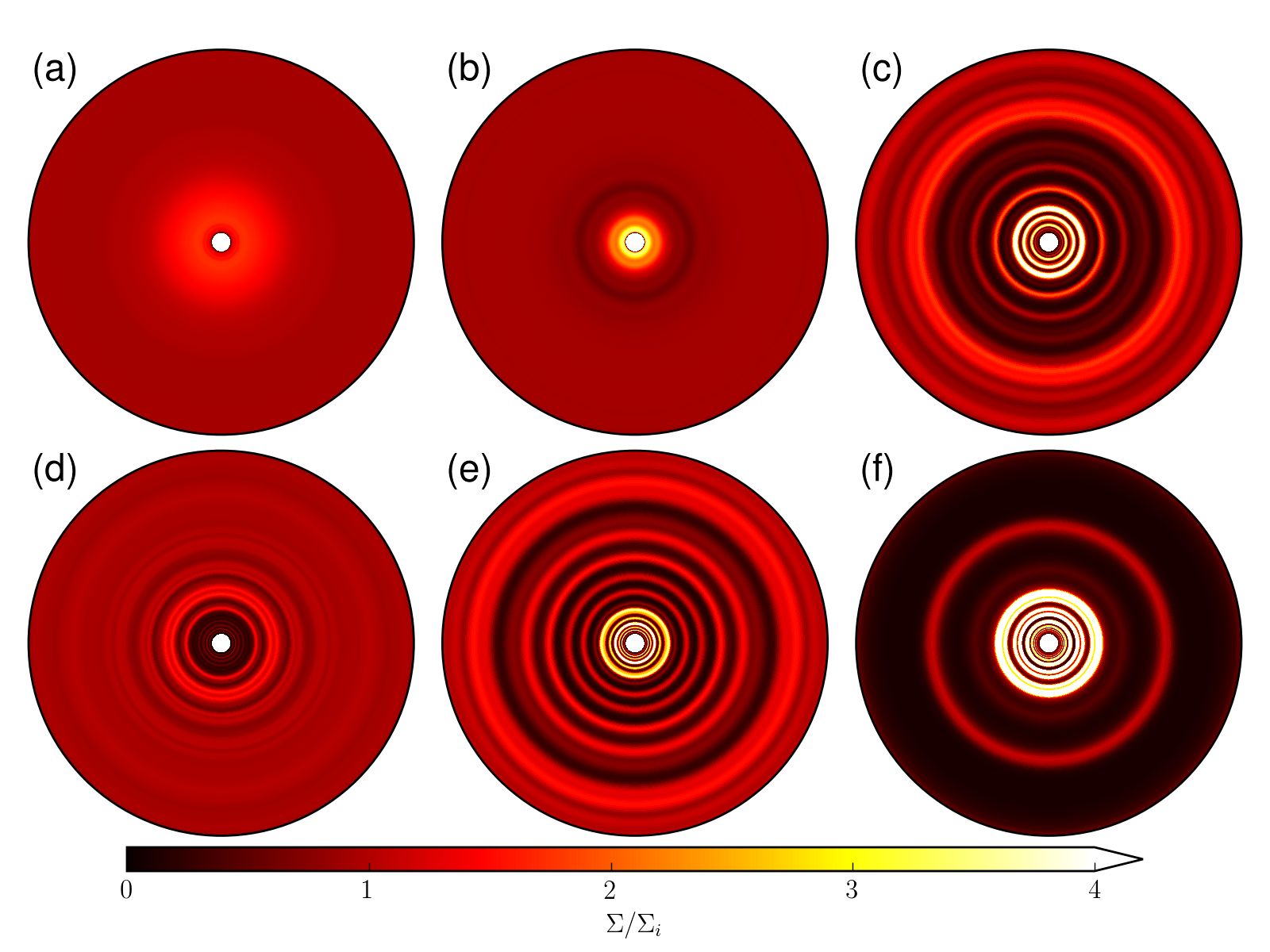}
    \caption{Face-on surface density profiles (up to a radius of 20~au) of the simulations where the explicit Ohmic resistivity and plasma-$\beta$ are varied at $t/t_0=2000$. In the top row, the explicit resistivity is decreased from panels (a)-(c). Plasma-$\beta$ varies from high to low across the bottom row in panels (d)-(f). The reference simulation (ad-els0.25) is shown in panel (e). The simulation panels in alphabetical order are: (a) oh26; (b) oh2.6; (c) oh0.26; (d) beta1e4; (e) ad-els0.25; (f) beta1e2. (See the supplementary material in the online journal for an animated version of this figure.)}
    \label{fig:face-on_beta}
\end{figure*}

\begin{figure}
    \centering
	\includegraphics[width=1.0\columnwidth]{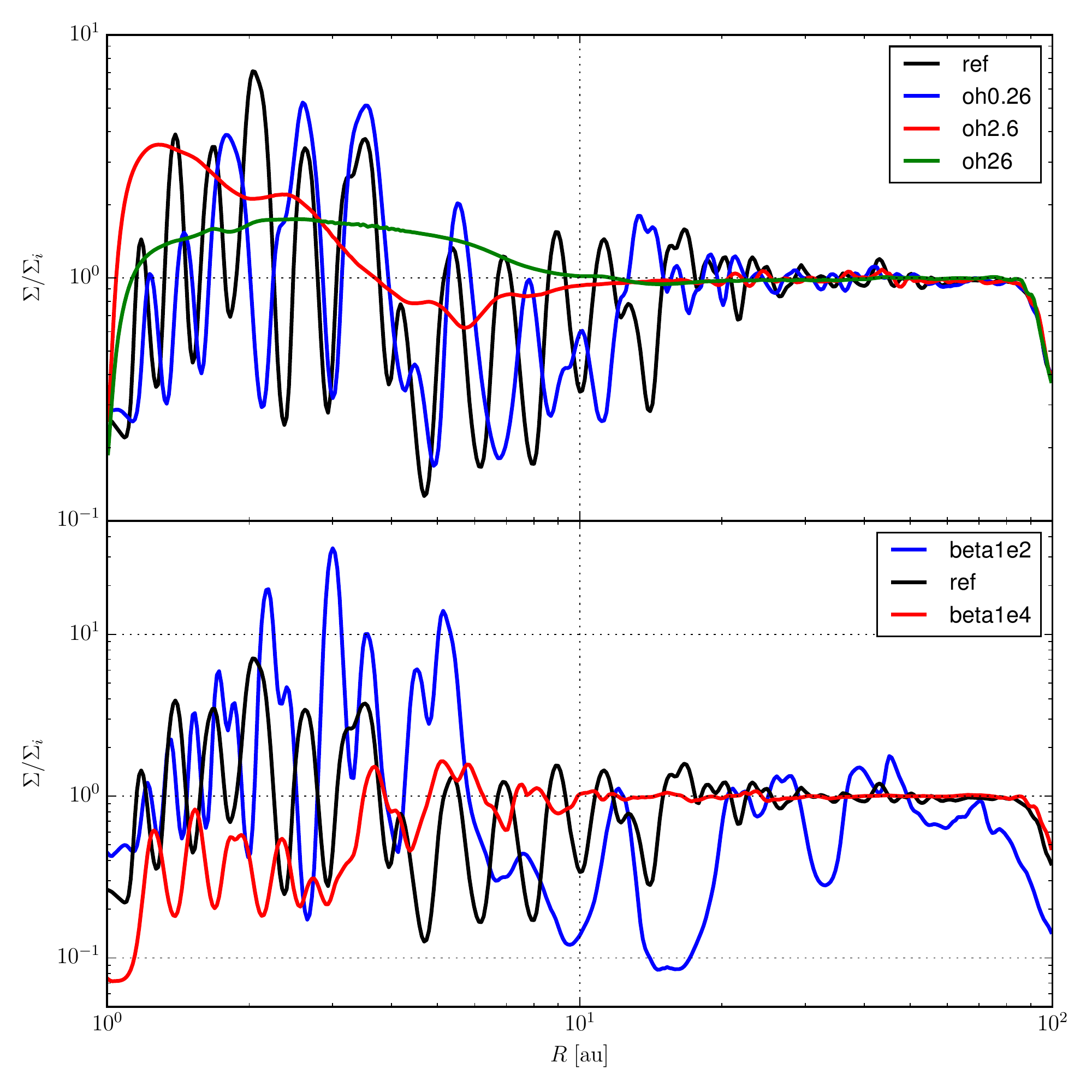}
    \caption{Surface density profiles at time $t/t_0=2000$ for simulations with different explicit Ohmic resistivities (top) and initial magnetic field strengths, i.e., $\beta_0$ (bottom). The surface density profiles are normalized to the initial surface density distribution, $\Sigma_i = \Sigma_0 (r/r_0)^{-1/2}$.}
    \label{fig:surfden1D_beta}
\end{figure}

\subsection{Explicit resistivity and magnetic field strength}\label{sec:beta}
The introduction of explicit Ohmic resistivity into the reference simulation can give us some important insights on the ring and gap formation mechanism. We add an Ohmic resistivity, $\eta_O$, that is constant in both space and time (as in \citetalias{2017MNRAS.468.3850S}). Specifically, three values of $\eta_0$ are considered corresponding to 0.26, 2.6, and 26 times the effective ambipolar resistivity at the inner edge of the disk at the midplane, $\eta_{A,0}=9.71\times 10^{14}$~cm$^2$s$^{-1}$; they are named oh0.26, oh2.6, and oh26, respectively. These simulations are plotted in panels (a)-(c) of Fig.~\ref{fig:beta_panels} and \ref{fig:face-on_beta}. In the most diffusive case with $\eta_O=26~\eta_{A,0}$, there is some concentration of mass at small radii, indicating that there is still mass accretion. However, there is little evidence for rings and gaps with the formation mechanism apparently turned off by the addition of a large resistivity. This strengthens the case for reconnection-driven ring and gap formation, because the large resistivity erases the sharp magnetic field geometries needed for reconnection. As the resistivity decreases, rings and gaps start to appear. In particular, when the resistivity $\eta_O$ drops below the characteristic AD resistivity $\eta_{A,0}$ (model oh0.26), the simulation looks very similar to the reference run that does not have any explicit resistivity. Their similarity, particularly in the location and structure of the rings and gaps, is quantified in Fig.~\ref{fig:surfden1D_beta}(a).

Besides magnetic diffusivity, the magnetic field strength also strongly affects the ring and gap formation. The second column of Fig.~\ref{fig:beta_panels} and \ref{fig:face-on_beta} (panels d-f) shows the effects of varying the initial magnetic field strength, as characterized by the midplane plasma-$\beta$, keeping everything else the same as in the reference run (panel e). These simulations are quantitatively similar, in that a wind is launched from the disk and rings and gaps are formed in all three cases. However, it takes longer for the weaker magnetic field case to produce a well-developed wind. Specifically, in the most weakly magnetized case (model beta1e4), it takes approximately 700 inner orbital periods for the disk wind from the inner part of the disk to become fully developed. This is because it takes longer to generate a strong enough toroidal field out of the weaker initial poloidal field to push the outer layers of the disk to large distances. The weakest field case should be the most prone to the MRI, however, there is no evidence for accretion streams developing near the disk surface.

As in the reference run, the ambipolar diffusion is able to concentrate the radial current ($J_r$) into a thin sheet near the midplane, where preferential accretion leads to severe radial pinching of the poloidal field, eventually leading to reconnection-driven ring and gap formation. The rings and gaps formed in this simulation have a relatively low contrast, however. This is because, with a weak initial field, there is less poloidal magnetic flux concentrated in the gaps after reconnection making the accretion of disk material from the gaps into the neighbouring rings less efficient.

In the stronger magnetic field case (model beta1e2), a quasi-steady wind is quickly established (see panel f of Fig.~\ref{fig:beta_panels}). It drives fast disk accretion, especially near the midplane, where reconnection of the sharply pinched poloidal field leads to demagnetization in some regions (creating rings) and bunching of poloidal field lines in others (creating gaps). The stronger poloidal field drives a more complete depletion of disk material, creating wider gaps with lower column densities, as illustrated in Fig.~\ref{fig:face-on_beta}(f) and quantified in Fig.~\ref{fig:surfden1D_beta}(b). In addition, the stronger overall field allows more material to be moved from the outer part of the disk to smaller disk radii, where several rings have much higher surface densities than their counterparts in the weaker field cases. The most massive inner ring at $r=3$~au has a contrast ratio of $\sim 10^2$. In any case, in the presence of the reference level of ambipolar diffusion, the same reconnection-driven ring and gap formation mechanism appears to operate over a range of disk plasma-$\beta$ with more strongly magnetized disks forming rings and gaps with higher surface density contrast.

\begin{figure}
    \centering
	\includegraphics[width=1.0\columnwidth]{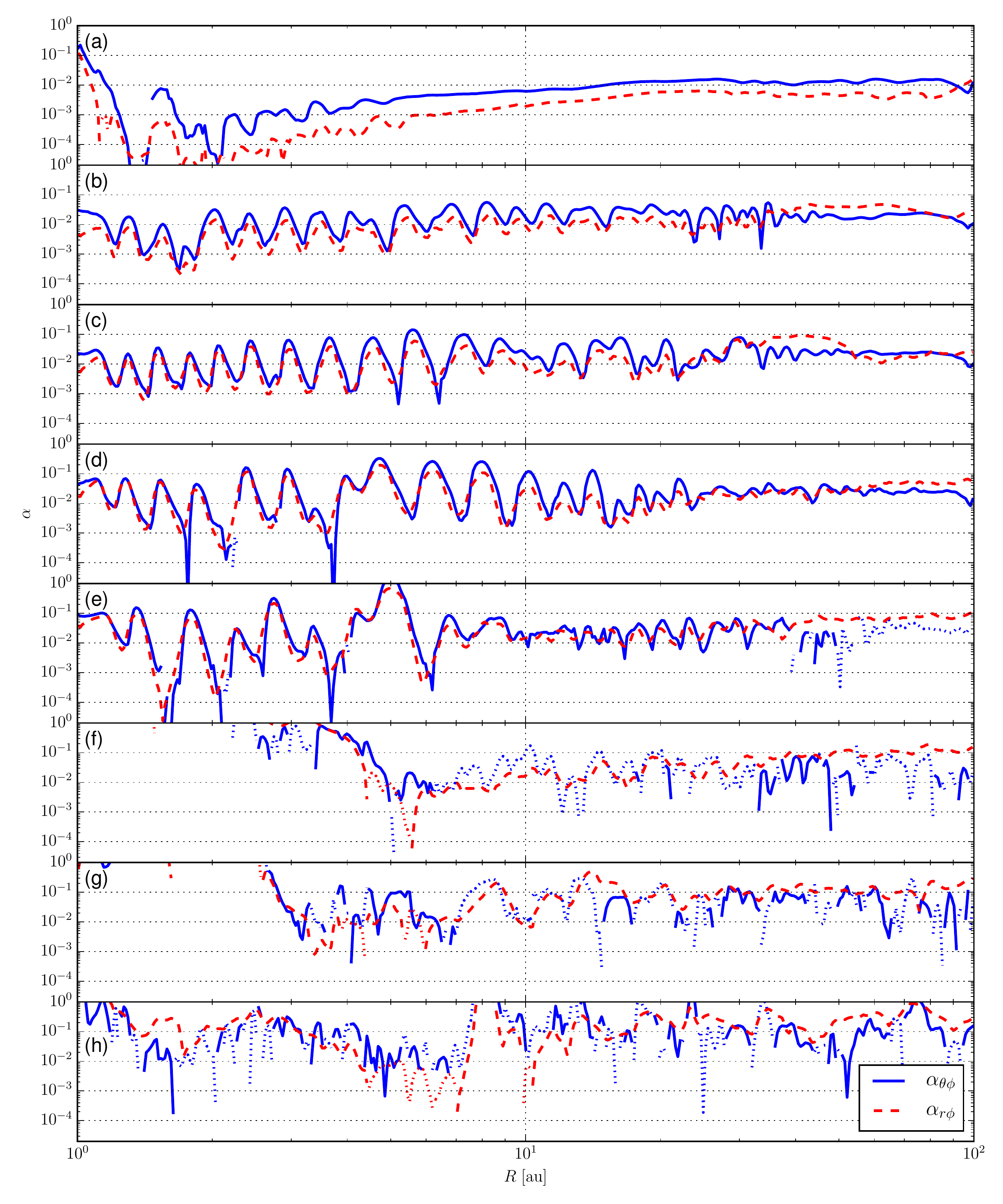}
    \caption{The $\alpha$ parameter from the vertical wind stress, $\alpha_{\theta\phi}$ (see equation~\ref{eq:alpha_theta}; solid blue line), and the radial shear stress, $\alpha_{r\phi}$ (see equation~\ref{eq:alpha_radial}; dashed red line). The vertical wind stress is calculated at the surface $\theta=\pi/2\pm2\epsilon$ and the radial shear stress is integrated between these surfaces. The dotted lines show where the effective $\alpha$ is negative. The AD Elsasser number increases sequentially from the top panel to the bottom panel: (a) ad-els0.01; (b) ad-els0.05; (c) ad-els0.1; (d) ad-els0.25 (ref); (e) ad-els0.5; (f) ad-els1.0; (g) ad-els2.0; (h) ideal.}
    \label{fig:stress}
\end{figure}

\subsection{Magnetic stresses and two modes of accretion}\label{sec:stress}
We have shown in Section~\ref{sec:els} and \ref{sec:beta} that the level of magnetic diffusivity, particularly ambipolar diffusion, plays a key role in determining the structure of the magnetically coupled disk-wind system. Specifically, more magnetically diffusive systems tend to be more laminar, with a well-developed wind that is expected to play a dominant role in driving disk accretion. Better magnetically coupled systems are more prone to MRI channel flows, which drive the system to a chaotic state. Although a wind is still developed, it may not play as important a role in disk accretion. In this subsection, we will try to quantify this expectation.

We do this through the dimensionless $\alpha$ parameters \citep{1973A&A....24..337S}, corresponding to the $r\phi$ and $z\phi$ components of the Maxwell stress, $T_{r\phi}=-\frac{B_r B_\phi}{4\pi}$ and $T_{\theta\phi}=-\frac{B_\theta B_\phi}{4\pi}$, respectively, defined as:
\begin{equation}
\alpha_{r\phi}\equiv  \frac{\int T_{r\phi} dz} {\int P dz},\label{eq:alpha_theta}
\end{equation}
\begin{equation}
\alpha_{\theta\phi}\equiv \frac{T_{\theta\phi}\Big\vert_{\pi/2-\theta_0}^{\pi/2+\theta_0}}{P_\mathrm{mid}}, \label{eq:alpha_radial}
\end{equation}
where $P$ is the thermal pressure and $P_\mathrm{mid}$ is the pressure on the disk midplane. In the first term, the integration is between the lower and upper surfaces of the initial disk, $\theta=\pi/2\pm\theta_0$. The second term is evaluated at these surfaces. Since the polar shear stress $T_{\theta\phi}$ is the magnetic stress that moves angular momentum across the disk surface (at a constant polar angle $\theta$), we will refer it to as the `wind stress.' The other component, $T_{r\phi}$, will be referred to as the `radial shear stress.'

Figure \ref{fig:stress} compares these two $\alpha$ parameters for the set of simulations where the AD Elsasser number is varied. In the least coupled simulation, the wind stress is larger than the radial shear stress by a factor of a few. As the Elsasser number increases, the radial shear stress begins to grow until it is approximately equal to the wind stress by the simulation where the AD Elsasser number is $\Lambda_0=0.5$ (see Fig.~\ref{fig:stress}e). In the intermediate parameter regime ($\Lambda_0$ between 0.05 and 0.5), where the wind remains relatively laminar and the rings and gaps are rather steady, the two stresses are strongly correlated with both peaking in low-density gaps where the poloidal magnetic field lines are concentrated (compare, for example, panel d of Fig.~\ref{fig:stress} to panel d of Fig.~\ref{fig:surfden1D}, which shows that both stresses peak in the low-density gaps for the reference run). As the Elsasser number increases further, the avalanche accretion streams become prevalent, driving the atmosphere of disk and the base of the wind to be chaotic. This transition to a more chaotic disk-wind system is already present in the outer part of the disk in the $\Lambda_0=0.5$ case beyond $\sim 50$~au (see panel e), where the avalanche accretion flows have reversed the direction of the polar shear stress $T_{\theta\phi}$ at the initial surface of the disk ($\theta=\pi/2\pm\theta_0$). For the cases with higher $\Lambda_0=1.0,~2.0,$ and $\infty$~(ideal MHD), the effective $\alpha$ parameter for $T_{\theta\phi}$ becomes highly variable in both space and time and is often negative. However, $T_{r\phi}$ stays mostly positive, indicating that angular momentum in the disk is more persistently transported radially outward by avalanche accretion streams rather than vertically  across the initial disk surface. 

\section{Discussion}\label{sec:discuss}

\subsection{Comparison to other works}
This work examines how radial substructure can be created in a circumstellar disk in the presence of ambipolar diffusion on the scale of a few to tens of au, as part of a magnetically coupled disk-wind system.  In \citetalias{2017MNRAS.468.3850S}, a similar phenomenon was observed to operate near the innermost disk radii ($\sim0.1$~au), where Ohmic resistivity dominates. As in \citetalias{2017MNRAS.468.3850S}, we again find that rings and gaps are formed solely from MHD processes. Here, the effects of AD have a clear and physically motivated interpretation as to how radial substructure is formed in simulations where the ions and neutrals are moderately coupled. The mechanism, described in Section~\ref{sec:ring}, relies on mass accretion through an AD-induced, midplane ($J_r$) current layer, where the poloidal magnetic field is dragged radially inward until it reconnects. The reconnection creates regions with magnetic loops where the net poloidal flux is decreased and mass accretion is less efficient, allowing matter to pile up into rings. It also enables the post-reconnection poloidal field to bunch up in localized regions, where mass accretion is more efficient, creating gaps.

The formation of radial disk substructures in MHD simulations (besides those formed by planets) has been seen at the boundaries of dead zones \citep{2010A&A...515A..70D,2015A&A...574A..68F,2016A&A...590A..17R} and in the context of zonal flows \citep{2009ApJ...697.1269J,2013MNRAS.434.2295K,2013ApJ...763..117D,2014ApJ...784...15S,2014ApJ...796...31B,2015ApJ...798...84B,2016A&A...589A..87B,2017A&A...600A..75B}. The concentration of poloidal magnetic field lines specifically in the presence of AD was observed in the disk simulations of \citet{2014ApJ...796...31B} and \citet{2017A&A...600A..75B}. This has been interpreted through the following form of the induction equation:
\begin{equation}
\frac{d\Phi_B(R,z)}{dt}=-2\pi R\mathcal{E}_\phi,
\end{equation}
where $\Phi_B$ is the vertical magnetic flux, and $\mathcal{E}_\phi$ is the $\phi$ component of the electromotive force (EMF; equation 8 of \citealt{2017ApJ...836...46B}; see also equation 23 of \citealt{2017A&A...600A..75B}). The EMF induced by AD is equal to $\bm{\mathcal{E}}_A=\eta_A \bm{J}_\perp$, where $\bm{J}_\perp$ is the component of the current perpendicular to the magnetic field (as defined in equation \ref{eq:induction}). When the azimuthal component of this perpendicular current, $J_{\perp,\phi}$, has a sign opposite to that of $J_\phi$, the AD EMF becomes anti-diffusive in nature, which would lead to the concentration of poloidal magnetic field lines. We have examined $J_{\perp,\phi}$ and $J_\phi$ in our reference run (where there is large spatial variation of the poloidal magnetic field strength, see Fig.~\ref{fig:bunch}b), and found that they have opposite signs in some regions but the same sign in others, which makes it hard to establish unambiguously the extent to which this mechanism may be operating in our simulations. In any case, we find that our results can be explained by a more pictorial mechanism: reconnection of sharply pinched poloidal field lines (e.g., Fig.~\ref{fig:reconnect}) that drives the segregation of poloidal magnetic flux relative to matter, which in turn leads to the formation of radial substructure. We note that \citet{2014ApJ...796...31B} also considered the possibility of reconnection playing a role in concentrating magnetic flux in the zonal flows found in their shearing box simulations (see their Fig.~9 for a cartoon illustrating the possibility). The relatively laminar nature of the disk accretion in the presence of a moderately strong ambipolar diffusion allowed us to isolate the reconnection events more clearly in our (2D) global simulations (see Fig.~\ref{fig:reconnect}). Whether it has a deeper physical connection with the mechanism that relies on $J_{\perp,\phi}$ and $J_\phi$ having opposite signs remains to be ascertained.

Although rings and gaps are prominent in most of our simulations, they are not a common feature of previous simulations. For example, the recent global accretion disk simulations of \citet{2017ApJ...836...46B} and \citet{2017ApJ...845...75B} do not seem to show such radial substructure. This is likely due to the fact that a weaker  initial poloidal field strength is used (however, see the shearing box simulations of \citealt{2015ApJ...798...84B} where zonal flows develop with $\beta=10^5$). Specifically, in the simulations of \citet{2017ApJ...845...75B} the initial magnetic field in the disk is characterized by $\beta=10^5$, which is higher than the largest initial value of plasma-$\beta$ used in our simulations. In our $\beta\sim10^4$ simulation, rings and gaps are still present. However, the surface density contrast is reduced compared to the reference run of $\beta\sim10^3$ (see Fig.~\ref{fig:surfden1D_beta}). Although the same magnetic field variations and midplane pinching still occur in the weaker magnetic field simulation of $\beta\sim10^4$ (see Fig.~\ref{fig:beta_panels}c), the magnetic field is less able to move matter around to form rings and gaps and the timescale for the magnetic field to dynamically influence the matter will be longer compared to the stronger field case. As such, we expect the formation of rings and gaps to become increasingly less efficient as the magnetic field strength is reduced towards the purely hydrodynamic limit. 

The initial field strengths in the ideal simulations of \citet{2017arXiv170104627Z} are similar to those adopted here ($\beta_0\sim10^3$). They show that most of the accretion occurs in a vertically extended disk `envelope,' with radial (as opposed to vertical) transport of angular momentum playing a dominant role in driving disk accretion. As discussed extensively in \citetalias{2017MNRAS.468.3850S}, this is consistent with the development of avalanche accretions streams as the Ohmic resistivity is reduced. It is also in agreement with the simulations in this work as we move towards the ideal MHD regime of large Elsasser numbers (see Fig.~\ref{fig:stress}). This agreement strengthens the case for the transition from a laminar disk-wind system to a more chaotic system dominated by the rapid formation and break up of accretion streams as the magnetic diffusivity (either Ohmic or ambipolar) is reduced.

\subsection{Dust dynamics and grain growth}
Wind-driven laminar disk accretion is an important feature of the moderately well magnetically coupled disks studied both in this paper (from AD) and in \citetalias{2017MNRAS.468.3850S} (due to Ohmic dissipation). There is some evidence that such a laminar accretion may be required for the HL Tau disk. As stressed by \citet{2016ApJ...816...25P}, there is tension between the small scale height of (sub)millimeter-emitting dust grains (inferred from the lack of azimuthal variation in the gap widths for the inclined HL disk, indicating strong dust settling) and the substantial ongoing mass accretion observed in the system, which, if driven by turbulence, would require a turbulence too strong to allow for the inferred degree of dust settling. This tension can be removed if the accretion is driven by ordered magnetic stresses rather than MRI-induced turbulence \citep{2017ApJ...845...31H}, as in our simulations with high to moderate levels of AD (such as the reference run), since dust grains can still settle to the midplane even with strong accretion. Furthermore, rings and gaps are naturally produced in these laminarly accreting disk-wind systems through the AD-aided magnetic reconnection; this mechanism can in principle produce the rings and gaps observed in the HL Tau disk. In practice, our  model parameters are chosen for the purposes of illustrating the basic principles of ring and gap formation in the presence of ambipolar diffusion rather than for comparison with any specific object. Taken at the face value, the typical mass accretion rate of $10^{-6}~\msunperyr$ found in the reference simulation is at least an order of magnitude larger than that inferred for classical T Tauri stars \citep{2016ARA&A..54..135H}. However, it is more consistent with the accretion rates inferred for younger protostellar disks (e.g., \citealt{2017ApJ...834..178Y}), although, it is possible to reduce the mass accretion rates in these simulations through rescaling (e.g., by adopting a lower initial disk density, $\rho_0$; see Appendix of \citealt{2014ApJ...793...31S}).    

The formation of rings and gaps in a relatively laminar disk has important implications for the dynamics of dust grains. Pressure maxima, such as those formed from dense gas rings, are known sites of dust trapping \citep{1972fpp..conf..211W,2010AREPS..38..493C}. Without such traps, large millimeter-sized grains would migrate inward quickly as they lose angular momentum to the more slowly rotating gas that is partially supported by the radial pressure gradient \citep{1977MNRAS.180...57W}. This rapid radial drift is particularly problematic for low-mass disks around brown dwarfs \citep{2013A&A...554A..95P}. For example, in the case of 2M0444, \citet{2017ApJ...846...19R} has shown explicitly that, without any dust trap, millimeter-sized grains would be quickly depleted from the outer part of this disk (on the scale of tens to a hundred au; see the upper-left panel of their Fig.~3), in direct contradiction to observations. They also demonstrated that this fundamental problem can be resolved if there are multiple pressure peaks in the outer disk (see the lower-right panel of their Fig.~3). Such pressure peaks are naturally produced in our simulations (see, e.g., Fig.~\ref{fig:bunch} of the reference run). 
 
Our mechanism of producing rings has two strengths. First, it takes into account ambipolar diffusion, which is the dominant non-ideal MHD effect in the outer disk where dust trapping is needed to be consistent with dust continuum observations. Second, it can in principle operate not only in relatively evolved protoplanetary disks but also younger protostellar disks as long as such disks are significantly magnetized with a poloidal field. Indeed, our mechanism is likely to work more efficiently in the earlier phases of disk evolution where the disk is expected to be threaded by a strong poloidal magnetic field, perhaps inherited from the collapse of dense cores, which are known to be magnetized with rather ordered magnetic fields (e.g., \citealt{2008ApJ...680..457T,2014prpl.conf..173L,2014prpl.conf..101L}). Such ordered poloidal fields can drive fast disk accretion expected in the early phases without generating a high level of turbulence in the outer (AD-dominated) region, which should make it easier for the dust to settle vertically and grow near the midplane, even during the early, perhaps Class 0, phase of star formation. In other words, strong accretion does not necessarily mean strong turbulence. Even in the earliest, Class 0 phase of star formation, large grains (if they are present) can be trapped in principle by the pressure bumps that naturally develop in the magnetically coupled disk-wind systems. Observationally, whether rings and gaps are prevalent in Class 0 disks is unknown at the present time, because they are more difficult to observe in the presence of a massive protostellar envelope, however, there is some evidence that rings and gaps are already present in at least the Class I phase (see observations of IRS 63 in $\rho$ Oph by Segura-Cox et al., \textit{in prep.}). 

Lastly, we note that the laminar disk wind in our reference and related simulations can preferentially remove gas from the disk, if the dust has settled to the midplane (or perhaps been trapped near the rings). As discussed by \citet{2010ApJ...718.1289S}, this could lead to an increase in the dust-to-gas mass ratio (see also \citealt{2015ApJ...804...29G,2016ApJ...818..152B}), conducive to the development of the streaming instability \citep{2005ApJ...620..459Y,2017arXiv171103975S}, which may facilitate the formation of planetesimals and eventually planets (e.g., \citealt{2010AREPS..38..493C}). This process of grain settling, growth, and trapping may be as efficient, in not more, in the early, protostellar phase of star formation compared to the later, protoplanetary phase.  We will postpone a quantitative exploration of this interesting topic to a future investigation.

\section{Conclusion}\label{sec:conc}

We have carried out 2D (axisymmetric) simulations of magnetically coupled disk-wind systems in the presence of a poloidal magnetic field and ambipolar diffusion (AD). The field strength is characterized by the plasma-$\beta$ and AD by the dimensionless Elsasser number $\Lambda_0$. We focused on $\beta\sim10^3$ and explored a wide range of values for $\Lambda_0$, from 0.01 to $\infty$ (ideal MHD). Our main conclusions are as follows: 
\begin{enumerate}
\item In moderately well coupled systems with $\Lambda_0$ between 0.05 and 0.5, including the reference simulation (ad-els0.25), we find that prominent rings and gaps are formed in the disk through a novel mechanism, AD-assisted reconnection. This mechanism starts with the twisting of the initial poloidal magnetic field into a toroidal field that reverses polarity across the disk midplane. Ambipolar diffusion enables the Lorentz force from the toroidal field pressure gradient to drive the ions (and the toroidal field lines tied to them) towards the magnetic null near the midplane, which steepens the radial ($J_r$) current sheet in a run-away process first described in \citet{1994ApJ...427L..91B}. The field kink generates a toroidal Lorentz force that removes angular momentum from the thin radial current sheet, forcing it to accrete preferentially relative to the rest of the disk. The preferential midplane accretion drags the poloidal field lines into a sharply pinched configuration, where the radial component of the magnetic field reverses polarity over a thin, secondary azimuthal ($J_\phi$) current sheet. Reconnection of the radial pinch produces two types of regions with distinct poloidal field topologies: one occupied by magnetic loops and another that remains threaded by ordered poloidal fields. The weakening of the net poloidal field in the former makes angular momentum removal less efficient, allowing disk material to accumulate to form dense rings. Conversely, those regions that gained poloidal flux after reconnection are magnetically braked more strongly, with a faster draining of disk material that leads to gap formation. In addition, AD allows for a quasi-steady state of the ring and gap structure, where the field lines can stay more or less fixed in place despite rapid mass accretion in gaps because of the ion-neutral drift. We find little evidence for the formation of prominent rings and gaps in the case of the highest ambipolar diffusion considered in this work ($\Lambda=0.01$) and cases with large, additional Ohmic diffusivities. This finding is consistent with the above scenario because the radial pinching of the poloidal field is smoothed out by the excessive magnetic diffusivity, suppressing the reconnection that lies at the heart of the mechanism.

\item In better magnetically coupled disk-wind systems with larger $\Lambda_0$, as well as the ideal MHD limit, we find that avalanche accretion streams develop spontaneously near the disk surface. The accretion streams lead to unsteady/chaotic disk accretion and outflow, as found previously in \citetalias{2017MNRAS.468.3850S} for cases of low or zero Ohmic resistivities (see also \citealt{2017arXiv170104627Z}). Prominent rings and gaps are still formed in the disk. Part of the reason is the large temporal and spatial variations induced by the constant formation and destruction of the streams will inevitably produce spatial variation in the mass accretion rate and thus the surface density. Perhaps more importantly, the poloidal field lines are concentrated in some regions and excluded from others, with the more strongly magnetized regions producing gaps and the less magnetized regions forming rings, just as in the more magnetically diffusive reference case. We suggest that this segregation of poloidal magnetic flux and matter is also due to reconnection of highly pinched poloidal fields. In this case, the pinching is caused by the avalanche accretion streams (a form of MRI channel flows) rather than the midplane current sheet steepened by AD. The fact that rings and gaps are formed in both laminar and chaotic disk-wind systems over a wide range of magnetic diffusivities suggests that they are a robust feature of such systems, at least when the initial poloidal magnetic field is relatively strong. For more weakly magnetized systems, reconnection may still occur but the resulting redistribution of poloidal magnetic flux would have less of a dynamical effect on the gas, making ring and gap formation less efficient.

\item If young star disks are threaded by a significant poloidal magnetic field, especially during the early phases of star formation, it may drive rapid disk accretion through a magnetic wind without necessarily generating strong turbulence in the disk, particularly in the outer parts of the disk that are only moderately well coupled to the magnetic field. The lack of a strong turbulence despite rapid accretion may allow dust to settle early in the process of star formation, facilitating early grain growth. Large grains may be trapped in the rings that are naturally produced in the system, which may promote the formation of planetesimals and eventually planets.
\end{enumerate}

\section*{Acknowledgements}
We thank Xuening Bai, Zhaohuan Zhu, and Takeru Suzuki for helpful discussions and the referee, William B\'{e}thune, for detailed constructive comments. This work is supported in part by National Science Foundation (NSF) grant AST-1313083 and AST-1716259 and National Aeronautics and Space Administration (NASA) grant NNX14AB38G.

\bibliography{main}{}

\begin{thebibliography}{}
\makeatletter
\relax
\def\mn@urlcharsother{\let\do\@makeother \do\$\do\&\do\#\do\^\do\_\do\%\do\~}
\def\mn@doi{\begingroup\mn@urlcharsother \@ifnextchar [ {\mn@doi@}
  {\mn@doi@[]}}
\def\mn@doi@[#1]#2{\def\@tempa{#1}\ifx\@tempa\@empty \href
  {http://dx.doi.org/#2} {doi:#2}\else \href {http://dx.doi.org/#2} {#1}\fi
  \endgroup}
\def\mn@eprint#1#2{\mn@eprint@#1:#2::\@nil}
\def\mn@eprint@arXiv#1{\href {http://arxiv.org/abs/#1} {{\tt arXiv:#1}}}
\def\mn@eprint@dblp#1{\href {http://dblp.uni-trier.de/rec/bibtex/#1.xml}
  {dblp:#1}}
\def\mn@eprint@#1:#2:#3:#4\@nil{\def\@tempa {#1}\def\@tempb {#2}\def\@tempc
  {#3}\ifx \@tempc \@empty \let \@tempc \@tempb \let \@tempb \@tempa \fi \ifx
  \@tempb \@empty \def\@tempb {arXiv}\fi \@ifundefined
  {mn@eprint@\@tempb}{\@tempb:\@tempc}{\expandafter \expandafter \csname
  mn@eprint@\@tempb\endcsname \expandafter{\@tempc}}}

\bibitem[\protect\citeauthoryear{{ALMA Partnership} et~al.,}{{ALMA Partnership}
  et~al.}{2015}]{2015ApJ...808L...3A}
{ALMA Partnership} et~al., 2015, \mn@doi [\apjl] {10.1088/2041-8205/808/1/L3},
  \href {http://adsabs.harvard.edu/abs/2015ApJ...808L...3A} {808, L3}

\bibitem[\protect\citeauthoryear{{Anderson}, {Li}, {Krasnopolsky}  \&
  {Blandford}}{{Anderson} et~al.}{2005}]{2005ApJ...630..945A}
{Anderson} J.~M.,  {Li} Z.-Y.,  {Krasnopolsky} R.,   {Blandford} R.~D.,  2005,
  \mn@doi [\apj] {10.1086/432040}, \href
  {http://adsabs.harvard.edu/abs/2005ApJ...630..945A} {630, 945}

\bibitem[\protect\citeauthoryear{{Andrews} et~al.,}{{Andrews}
  et~al.}{2016}]{2016ApJ...820L..40A}
{Andrews} S.~M.,  et~al., 2016, \mn@doi [\apjl] {10.3847/2041-8205/820/2/L40},
  \href {http://adsabs.harvard.edu/abs/2016ApJ...820L..40A} {820, L40}

\bibitem[\protect\citeauthoryear{{Bae}, {Zhu}  \& {Hartmann}}{{Bae}
  et~al.}{2017}]{2017ApJ...850..201B}
{Bae} J.,  {Zhu} Z.,   {Hartmann} L.,  2017, \mn@doi [\apj]
  {10.3847/1538-4357/aa9705}, \href
  {http://adsabs.harvard.edu/abs/2017ApJ...850..201B} {850, 201}

\bibitem[\protect\citeauthoryear{{Bai}}{{Bai}}{2013}]{2013ApJ...772...96B}
{Bai} X.-N.,  2013, \mn@doi [\apj] {10.1088/0004-637X/772/2/96}, \href
  {http://adsabs.harvard.edu/abs/2013ApJ...772...96B} {772, 96}

\bibitem[\protect\citeauthoryear{{Bai}}{{Bai}}{2015}]{2015ApJ...798...84B}
{Bai} X.-N.,  2015, \mn@doi [\apj] {10.1088/0004-637X/798/2/84}, \href
  {http://adsabs.harvard.edu/abs/2015ApJ...798...84B} {798, 84}

\bibitem[\protect\citeauthoryear{{Bai}}{{Bai}}{2017}]{2017ApJ...845...75B}
{Bai} X.-N.,  2017, \mn@doi [\apj] {10.3847/1538-4357/aa7dda}, \href
  {http://adsabs.harvard.edu/abs/2017ApJ...845...75B} {845, 75}

\bibitem[\protect\citeauthoryear{{Bai} \& {Goodman}}{{Bai} \&
  {Goodman}}{2009}]{2009ApJ...701..737B}
{Bai} X.-N.,  {Goodman} J.,  2009, \mn@doi [\apj]
  {10.1088/0004-637X/701/1/737}, \href
  {http://adsabs.harvard.edu/abs/2009ApJ...701..737B} {701, 737}

\bibitem[\protect\citeauthoryear{{Bai} \& {Stone}}{{Bai} \&
  {Stone}}{2011}]{2011ApJ...736..144B}
{Bai} X.-N.,  {Stone} J.~M.,  2011, \mn@doi [\apj]
  {10.1088/0004-637X/736/2/144}, \href
  {http://adsabs.harvard.edu/abs/2011ApJ...736..144B} {736, 144}

\bibitem[\protect\citeauthoryear{{Bai} \& {Stone}}{{Bai} \&
  {Stone}}{2013}]{2013ApJ...769...76B}
{Bai} X.-N.,  {Stone} J.~M.,  2013, \mn@doi [\apj]
  {10.1088/0004-637X/769/1/76}, \href
  {http://adsabs.harvard.edu/abs/2013ApJ...769...76B} {769, 76}

\bibitem[\protect\citeauthoryear{{Bai} \& {Stone}}{{Bai} \&
  {Stone}}{2014}]{2014ApJ...796...31B}
{Bai} X.-N.,  {Stone} J.~M.,  2014, \mn@doi [\apj]
  {10.1088/0004-637X/796/1/31}, \href
  {http://adsabs.harvard.edu/abs/2014ApJ...796...31B} {796, 31}

\bibitem[\protect\citeauthoryear{{Bai} \& {Stone}}{{Bai} \&
  {Stone}}{2017}]{2017ApJ...836...46B}
{Bai} X.-N.,  {Stone} J.~M.,  2017, \mn@doi [\apj]
  {10.3847/1538-4357/836/1/46}, \href
  {http://adsabs.harvard.edu/abs/2017ApJ...836...46B} {836, 46}

\bibitem[\protect\citeauthoryear{{Bai}, {Ye}, {Goodman}  \& {Yuan}}{{Bai}
  et~al.}{2016}]{2016ApJ...818..152B}
{Bai} X.-N.,  {Ye} J.,  {Goodman} J.,   {Yuan} F.,  2016, \mn@doi [\apj]
  {10.3847/0004-637X/818/2/152}, \href
  {http://adsabs.harvard.edu/abs/2016ApJ...818..152B} {818, 152}

\bibitem[\protect\citeauthoryear{{Balbus} \& {Hawley}}{{Balbus} \&
  {Hawley}}{1991}]{1991ApJ...376..214B}
{Balbus} S.~A.,  {Hawley} J.~F.,  1991, \mn@doi [\apj] {10.1086/170270}, \href
  {http://adsabs.harvard.edu/abs/1991ApJ...376..214B} {376, 214}

\bibitem[\protect\citeauthoryear{{B{\'e}thune}, {Lesur}  \&
  {Ferreira}}{{B{\'e}thune} et~al.}{2016}]{2016A&A...589A..87B}
{B{\'e}thune} W.,  {Lesur} G.,   {Ferreira} J.,  2016, \mn@doi [\aap]
  {10.1051/0004-6361/201527874}, \href
  {http://adsabs.harvard.edu/abs/2016A%26A...589A..87B} {589, A87}

\bibitem[\protect\citeauthoryear{{B{\'e}thune}, {Lesur}  \&
  {Ferreira}}{{B{\'e}thune} et~al.}{2017}]{2017A&A...600A..75B}
{B{\'e}thune} W.,  {Lesur} G.,   {Ferreira} J.,  2017, \mn@doi [\aap]
  {10.1051/0004-6361/201630056}, \href
  {http://adsabs.harvard.edu/abs/2017A%26A...600A..75B} {600, A75}

\bibitem[\protect\citeauthoryear{{Blandford} \& {Payne}}{{Blandford} \&
  {Payne}}{1982}]{1982MNRAS.199..883B}
{Blandford} R.~D.,  {Payne} D.~G.,  1982, \mn@doi [\mnras]
  {10.1093/mnras/199.4.883}, \href
  {http://adsabs.harvard.edu/abs/1982MNRAS.199..883B} {199, 883}

\bibitem[\protect\citeauthoryear{{Brandenburg} \& {Zweibel}}{{Brandenburg} \&
  {Zweibel}}{1994}]{1994ApJ...427L..91B}
{Brandenburg} A.,  {Zweibel} E.~G.,  1994, \mn@doi [\apjl] {10.1086/187372},
  \href {http://adsabs.harvard.edu/abs/1994ApJ...427L..91B} {427, L91}

\bibitem[\protect\citeauthoryear{{Casse} \& {Ferreira}}{{Casse} \&
  {Ferreira}}{2000}]{2000A&A...353.1115C}
{Casse} F.,  {Ferreira} J.,  2000, \aap, \href
  {http://adsabs.harvard.edu/abs/2000A%26A...353.1115C} {353, 1115}

\bibitem[\protect\citeauthoryear{{Chiang} \& {Murray-Clay}}{{Chiang} \&
  {Murray-Clay}}{2007}]{2007NatPh...3..604C}
{Chiang} E.,  {Murray-Clay} R.,  2007, \mn@doi [Nature Physics]
  {10.1038/nphys661}, \href {http://adsabs.harvard.edu/abs/2007NatPh...3..604C}
  {3, 604}

\bibitem[\protect\citeauthoryear{{Chiang} \& {Youdin}}{{Chiang} \&
  {Youdin}}{2010}]{2010AREPS..38..493C}
{Chiang} E.,  {Youdin} A.~N.,  2010, \mn@doi [Annual Review of Earth and
  Planetary Sciences] {10.1146/annurev-earth-040809-152513}, \href
  {http://adsabs.harvard.edu/abs/2010AREPS..38..493C} {38, 493}

\bibitem[\protect\citeauthoryear{{Cieza} et~al.,}{{Cieza}
  et~al.}{2016}]{2016Natur.535..258C}
{Cieza} L.~A.,  et~al., 2016, \mn@doi [\nat] {10.1038/nature18612}, \href
  {http://adsabs.harvard.edu/abs/2016Natur.535..258C} {535, 258}

\bibitem[\protect\citeauthoryear{{Clarke}}{{Clarke}}{1996}]{1996ApJ...457..291C}
{Clarke} D.~A.,  1996, \mn@doi [\apj] {10.1086/176730}, \href
  {http://adsabs.harvard.edu/abs/1996ApJ...457..291C} {457, 291}

\bibitem[\protect\citeauthoryear{{Clarke}}{{Clarke}}{2010}]{2010ApJS..187..119C}
{Clarke} D.~A.,  2010, \mn@doi [\apjs] {10.1088/0067-0049/187/1/119}, \href
  {http://adsabs.harvard.edu/abs/2010ApJS..187..119C} {187, 119}

\bibitem[\protect\citeauthoryear{{Dipierro}, {Price}, {Laibe}, {Hirsh},
  {Cerioli}  \& {Lodato}}{{Dipierro} et~al.}{2015}]{2015MNRAS.453L..73D}
{Dipierro} G.,  {Price} D.,  {Laibe} G.,  {Hirsh} K.,  {Cerioli} A.,   {Lodato}
  G.,  2015, \mn@doi [\mnras] {10.1093/mnrasl/slv105}, \href
  {http://adsabs.harvard.edu/abs/2015MNRAS.453L..73D} {453, L73}

\bibitem[\protect\citeauthoryear{{Dittrich}, {Klahr}  \& {Johansen}}{{Dittrich}
  et~al.}{2013}]{2013ApJ...763..117D}
{Dittrich} K.,  {Klahr} H.,   {Johansen} A.,  2013, \mn@doi [\apj]
  {10.1088/0004-637X/763/2/117}, \href
  {http://adsabs.harvard.edu/abs/2013ApJ...763..117D} {763, 117}

\bibitem[\protect\citeauthoryear{{Dong}, {Zhu}  \& {Whitney}}{{Dong}
  et~al.}{2015}]{2015ApJ...809...93D}
{Dong} R.,  {Zhu} Z.,   {Whitney} B.,  2015, \mn@doi [\apj]
  {10.1088/0004-637X/809/1/93}, \href
  {http://adsabs.harvard.edu/abs/2015ApJ...809...93D} {809, 93}

\bibitem[\protect\citeauthoryear{{Dong}, {Li}, {Chiang}  \& {Li}}{{Dong}
  et~al.}{2017}]{2017ApJ...843..127D}
{Dong} R.,  {Li} S.,  {Chiang} E.,   {Li} H.,  2017, \mn@doi [\apj]
  {10.3847/1538-4357/aa72f2}, \href
  {http://adsabs.harvard.edu/abs/2017ApJ...843..127D} {843, 127}

\bibitem[\protect\citeauthoryear{{Dzyurkevich}, {Flock}, {Turner}, {Klahr}  \&
  {Henning}}{{Dzyurkevich} et~al.}{2010}]{2010A&A...515A..70D}
{Dzyurkevich} N.,  {Flock} M.,  {Turner} N.~J.,  {Klahr} H.,   {Henning} T.,
  2010, \mn@doi [\aap] {10.1051/0004-6361/200912834}, \href
  {http://adsabs.harvard.edu/abs/2010A%26A...515A..70D} {515, A70}

\bibitem[\protect\citeauthoryear{{Fedele} et~al.,}{{Fedele}
  et~al.}{2017}]{2017A&A...600A..72F}
{Fedele} D.,  et~al., 2017, \mn@doi [\aap] {10.1051/0004-6361/201629860}, \href
  {http://adsabs.harvard.edu/abs/2017A%26A...600A..72F} {600, A72}

\bibitem[\protect\citeauthoryear{{Fleming} \& {Stone}}{{Fleming} \&
  {Stone}}{2003}]{2003ApJ...585..908F}
{Fleming} T.,  {Stone} J.~M.,  2003, \mn@doi [\apj] {10.1086/345848}, \href
  {http://adsabs.harvard.edu/abs/2003ApJ...585..908F} {585, 908}

\bibitem[\protect\citeauthoryear{{Fleming}, {Stone}  \& {Hawley}}{{Fleming}
  et~al.}{2000}]{2000ApJ...530..464F}
{Fleming} T.~P.,  {Stone} J.~M.,   {Hawley} J.~F.,  2000, \mn@doi [\apj]
  {10.1086/308338}, \href {http://adsabs.harvard.edu/abs/2000ApJ...530..464F}
  {530, 464}

\bibitem[\protect\citeauthoryear{{Flock}, {Ruge}, {Dzyurkevich}, {Henning},
  {Klahr}  \& {Wolf}}{{Flock} et~al.}{2015}]{2015A&A...574A..68F}
{Flock} M.,  {Ruge} J.~P.,  {Dzyurkevich} N.,  {Henning} T.,  {Klahr} H.,
  {Wolf} S.,  2015, \mn@doi [\aap] {10.1051/0004-6361/201424693}, \href
  {http://adsabs.harvard.edu/abs/2015A%26A...574A..68F} {574, A68}

\bibitem[\protect\citeauthoryear{{Furth}, {Killeen}  \& {Rosenbluth}}{{Furth}
  et~al.}{1963}]{1963PhFl....6..459F}
{Furth} H.~P.,  {Killeen} J.,   {Rosenbluth} M.~N.,  1963, \mn@doi [Physics of
  Fluids] {10.1063/1.1706761}, \href
  {http://adsabs.harvard.edu/abs/1963PhFl....6..459F} {6, 459}

\bibitem[\protect\citeauthoryear{{Glassgold}, {Lizano}  \& {Galli}}{{Glassgold}
  et~al.}{2017}]{2017MNRAS.472.2447G}
{Glassgold} A.~E.,  {Lizano} S.,   {Galli} D.,  2017, \mn@doi [\mnras]
  {10.1093/mnras/stx2145}, \href
  {http://adsabs.harvard.edu/abs/2017MNRAS.472.2447G} {472, 2447}

\bibitem[\protect\citeauthoryear{{Gorti}, {Hollenbach}  \& {Dullemond}}{{Gorti}
  et~al.}{2015}]{2015ApJ...804...29G}
{Gorti} U.,  {Hollenbach} D.,   {Dullemond} C.~P.,  2015, \mn@doi [\apj]
  {10.1088/0004-637X/804/1/29}, \href
  {http://adsabs.harvard.edu/abs/2015ApJ...804...29G} {804, 29}

\bibitem[\protect\citeauthoryear{{Gressel}, {Turner}, {Nelson}  \&
  {McNally}}{{Gressel} et~al.}{2015}]{2015ApJ...801...84G}
{Gressel} O.,  {Turner} N.~J.,  {Nelson} R.~P.,   {McNally} C.~P.,  2015,
  \mn@doi [\apj] {10.1088/0004-637X/801/2/84}, \href
  {http://adsabs.harvard.edu/abs/2015ApJ...801...84G} {801, 84}

\bibitem[\protect\citeauthoryear{{Hartmann}, {Herczeg}  \& {Calvet}}{{Hartmann}
  et~al.}{2016}]{2016ARA&A..54..135H}
{Hartmann} L.,  {Herczeg} G.,   {Calvet} N.,  2016, \mn@doi [\araa]
  {10.1146/annurev-astro-081915-023347}, \href
  {http://adsabs.harvard.edu/abs/2016ARA%26A..54..135H} {54, 135}

\bibitem[\protect\citeauthoryear{{Hasegawa}, {Okuzumi}, {Flock}  \&
  {Turner}}{{Hasegawa} et~al.}{2017}]{2017ApJ...845...31H}
{Hasegawa} Y.,  {Okuzumi} S.,  {Flock} M.,   {Turner} N.~J.,  2017, \mn@doi
  [\apj] {10.3847/1538-4357/aa7d55}, \href
  {http://adsabs.harvard.edu/abs/2017ApJ...845...31H} {845, 31}

\bibitem[\protect\citeauthoryear{{Hirota}, {Machida}, {Matsushita}, {Motogi},
  {Matsumoto}, {Kim}, {Burns}  \& {Honma}}{{Hirota}
  et~al.}{2017}]{2017NatAs...1E.146H}
{Hirota} T.,  {Machida} M.~N.,  {Matsushita} Y.,  {Motogi} K.,  {Matsumoto} N.,
   {Kim} M.~K.,  {Burns} R.~A.,   {Honma} M.,  2017, \mn@doi [Nature Astronomy]
  {10.1038/s41550-017-0146}, \href
  {http://adsabs.harvard.edu/abs/2017NatAs...1E.146H} {1, 0146}

\bibitem[\protect\citeauthoryear{{Isella} et~al.,}{{Isella}
  et~al.}{2016}]{2016PhRvL.117y1101I}
{Isella} A.,  et~al., 2016, \mn@doi [Physical Review Letters]
  {10.1103/PhysRevLett.117.251101}, \href
  {http://adsabs.harvard.edu/abs/2016PhRvL.117y1101I} {117, 251101}

\bibitem[\protect\citeauthoryear{{Johansen}, {Youdin}  \& {Klahr}}{{Johansen}
  et~al.}{2009}]{2009ApJ...697.1269J}
{Johansen} A.,  {Youdin} A.,   {Klahr} H.,  2009, \mn@doi [\apj]
  {10.1088/0004-637X/697/2/1269}, \href
  {http://adsabs.harvard.edu/abs/2009ApJ...697.1269J} {697, 1269}

\bibitem[\protect\citeauthoryear{{Krasnopolsky}, {Li}  \&
  {Shang}}{{Krasnopolsky} et~al.}{2010}]{2010ApJ...716.1541K}
{Krasnopolsky} R.,  {Li} Z.-Y.,   {Shang} H.,  2010, \mn@doi [\apj]
  {10.1088/0004-637X/716/2/1541}, \href
  {http://adsabs.harvard.edu/abs/2010ApJ...716.1541K} {716, 1541}

\bibitem[\protect\citeauthoryear{{Kudoh}, {Matsumoto}  \& {Shibata}}{{Kudoh}
  et~al.}{1998}]{1998ApJ...508..186K}
{Kudoh} T.,  {Matsumoto} R.,   {Shibata} K.,  1998, \mn@doi [\apj]
  {10.1086/306377}, \href {http://adsabs.harvard.edu/abs/1998ApJ...508..186K}
  {508, 186}

\bibitem[\protect\citeauthoryear{{Kunz} \& {Lesur}}{{Kunz} \&
  {Lesur}}{2013}]{2013MNRAS.434.2295K}
{Kunz} M.~W.,  {Lesur} G.,  2013, \mn@doi [\mnras] {10.1093/mnras/stt1171},
  \href {http://adsabs.harvard.edu/abs/2013MNRAS.434.2295K} {434, 2295}

\bibitem[\protect\citeauthoryear{{Latter}, {Lesaffre}  \& {Balbus}}{{Latter}
  et~al.}{2009}]{2009MNRAS.394..715L}
{Latter} H.~N.,  {Lesaffre} P.,   {Balbus} S.~A.,  2009, \mn@doi [\mnras]
  {10.1111/j.1365-2966.2009.14395.x}, \href
  {http://adsabs.harvard.edu/abs/2009MNRAS.394..715L} {394, 715}

\bibitem[\protect\citeauthoryear{{Lee} et~al.,}{{Lee}
  et~al.}{2018}]{2018arXiv180203668L}
{Lee} C.-F.,  et~al., 2018, preprint, \href
  {http://adsabs.harvard.edu/abs/2018arXiv180203668L} {} (\mn@eprint {arXiv}
  {1802.03668})

\bibitem[\protect\citeauthoryear{{Lesur}, {Kunz}  \& {Fromang}}{{Lesur}
  et~al.}{2014}]{2014A&A...566A..56L}
{Lesur} G.,  {Kunz} M.~W.,   {Fromang} S.,  2014, \mn@doi [\aap]
  {10.1051/0004-6361/201423660}, \href
  {http://adsabs.harvard.edu/abs/2014A%26A...566A..56L} {566, A56}

\bibitem[\protect\citeauthoryear{{Li} \& {McKee}}{{Li} \&
  {McKee}}{1996}]{1996ApJ...464..373L}
{Li} Z.-Y.,  {McKee} C.~F.,  1996, \mn@doi [\apj] {10.1086/177329}, \href
  {http://adsabs.harvard.edu/abs/1996ApJ...464..373L} {464, 373}

\bibitem[\protect\citeauthoryear{{Li}, {Krasnopolsky}  \& {Shang}}{{Li}
  et~al.}{2011}]{2011ApJ...738..180L}
{Li} Z.-Y.,  {Krasnopolsky} R.,   {Shang} H.,  2011, \mn@doi [\apj]
  {10.1088/0004-637X/738/2/180}, \href
  {http://adsabs.harvard.edu/abs/2011ApJ...738..180L} {738, 180}

\bibitem[\protect\citeauthoryear{{Li}, {Goodman}, {Sridharan}, {Houde}, {Li},
  {Novak}  \& {Tang}}{{Li} et~al.}{2014a}]{2014prpl.conf..101L}
{Li} H.-B.,  {Goodman} A.,  {Sridharan} T.~K.,  {Houde} M.,  {Li} Z.-Y.,
  {Novak} G.,   {Tang} K.~S.,  2014a, \mn@doi [Protostars and Planets VI]
  {10.2458/azu_uapress_9780816531240-ch005}, \href
  {http://adsabs.harvard.edu/abs/2014prpl.conf..101L} {pp 101--123}

\bibitem[\protect\citeauthoryear{{Li}, {Banerjee}, {Pudritz}, {J{\o}rgensen},
  {Shang}, {Krasnopolsky}  \& {Maury}}{{Li}
  et~al.}{2014b}]{2014prpl.conf..173L}
{Li} Z.-Y.,  {Banerjee} R.,  {Pudritz} R.~E.,  {J{\o}rgensen} J.~K.,  {Shang}
  H.,  {Krasnopolsky} R.,   {Maury} A.,  2014b, \mn@doi [Protostars and Planets
  VI] {10.2458/azu_uapress_9780816531240-ch008}, \href
  {http://adsabs.harvard.edu/abs/2014prpl.conf..173L} {pp 173--194}

\bibitem[\protect\citeauthoryear{{Mac Low}, {Norman}, {Konigl}  \&
  {Wardle}}{{Mac Low} et~al.}{1995}]{1995ApJ...442..726M}
{Mac Low} M.-M.,  {Norman} M.~L.,  {Konigl} A.,   {Wardle} M.,  1995, \mn@doi
  [\apj] {10.1086/175477}, \href
  {http://adsabs.harvard.edu/abs/1995ApJ...442..726M} {442, 726}

\bibitem[\protect\citeauthoryear{{Matsumoto}, {Uchida}, {Hirose}, {Shibata},
  {Hayashi}, {Ferrari}, {Bodo}  \& {Norman}}{{Matsumoto}
  et~al.}{1996}]{1996ApJ...461..115M}
{Matsumoto} R.,  {Uchida} Y.,  {Hirose} S.,  {Shibata} K.,  {Hayashi} M.~R.,
  {Ferrari} A.,  {Bodo} G.,   {Norman} C.,  1996, \mn@doi [\apj]
  {10.1086/177041}, \href {http://adsabs.harvard.edu/abs/1996ApJ...461..115M}
  {461, 115}

\bibitem[\protect\citeauthoryear{{Moll}}{{Moll}}{2012}]{2012A&A...548A..76M}
{Moll} R.,  2012, \mn@doi [\aap] {10.1051/0004-6361/201118249}, \href
  {http://adsabs.harvard.edu/abs/2012A%26A...548A..76M} {548, A76}

\bibitem[\protect\citeauthoryear{{Nomura} et~al.,}{{Nomura}
  et~al.}{2016}]{2016ApJ...819L...7N}
{Nomura} H.,  et~al., 2016, \mn@doi [\apjl] {10.3847/2041-8205/819/1/L7}, \href
  {http://adsabs.harvard.edu/abs/2016ApJ...819L...7N} {819, L7}

\bibitem[\protect\citeauthoryear{{Okuzumi}, {Momose}, {Sirono}, {Kobayashi}  \&
  {Tanaka}}{{Okuzumi} et~al.}{2016}]{2016ApJ...821...82O}
{Okuzumi} S.,  {Momose} M.,  {Sirono} S.-i.,  {Kobayashi} H.,   {Tanaka} H.,
  2016, \mn@doi [\apj] {10.3847/0004-637X/821/2/82}, \href
  {http://adsabs.harvard.edu/abs/2016ApJ...821...82O} {821, 82}

\bibitem[\protect\citeauthoryear{{Perez-Becker} \& {Chiang}}{{Perez-Becker} \&
  {Chiang}}{2011a}]{2011ApJ...727....2P}
{Perez-Becker} D.,  {Chiang} E.,  2011a, \mn@doi [\apj]
  {10.1088/0004-637X/727/1/2}, \href
  {http://adsabs.harvard.edu/abs/2011ApJ...727....2P} {727, 2}

\bibitem[\protect\citeauthoryear{{Perez-Becker} \& {Chiang}}{{Perez-Becker} \&
  {Chiang}}{2011b}]{2011ApJ...735....8P}
{Perez-Becker} D.,  {Chiang} E.,  2011b, \mn@doi [\apj]
  {10.1088/0004-637X/735/1/8}, \href
  {http://adsabs.harvard.edu/abs/2011ApJ...735....8P} {735, 8}

\bibitem[\protect\citeauthoryear{{P{\'e}rez} et~al.,}{{P{\'e}rez}
  et~al.}{2016}]{2016Sci...353.1519P}
{P{\'e}rez} L.~M.,  et~al., 2016, \mn@doi [Science] {10.1126/science.aaf8296},
  \href {http://adsabs.harvard.edu/abs/2016Sci...353.1519P} {353, 1519}

\bibitem[\protect\citeauthoryear{{Pinilla}, {Birnstiel}, {Benisty}, {Ricci},
  {Natta}, {Dullemond}, {Dominik}  \& {Testi}}{{Pinilla}
  et~al.}{2013}]{2013A&A...554A..95P}
{Pinilla} P.,  {Birnstiel} T.,  {Benisty} M.,  {Ricci} L.,  {Natta} A.,
  {Dullemond} C.~P.,  {Dominik} C.,   {Testi} L.,  2013, \mn@doi [\aap]
  {10.1051/0004-6361/201220875}, \href
  {http://adsabs.harvard.edu/abs/2013A%26A...554A..95P} {554, A95}

\bibitem[\protect\citeauthoryear{{Pinte}, {Dent}, {M{\'e}nard}, {Hales},
  {Hill}, {Cortes}  \& {de Gregorio-Monsalvo}}{{Pinte}
  et~al.}{2016}]{2016ApJ...816...25P}
{Pinte} C.,  {Dent} W.~R.~F.,  {M{\'e}nard} F.,  {Hales} A.,  {Hill} T.,
  {Cortes} P.,   {de Gregorio-Monsalvo} I.,  2016, \mn@doi [\apj]
  {10.3847/0004-637X/816/1/25}, \href
  {http://adsabs.harvard.edu/abs/2016ApJ...816...25P} {816, 25}

\bibitem[\protect\citeauthoryear{{Ricci}, {Rome}, {Pinilla}, {Facchini},
  {Birnstiel}  \& {Testi}}{{Ricci} et~al.}{2017}]{2017ApJ...846...19R}
{Ricci} L.,  {Rome} H.,  {Pinilla} P.,  {Facchini} S.,  {Birnstiel} T.,
  {Testi} L.,  2017, \mn@doi [\apj] {10.3847/1538-4357/aa81bf}, \href
  {http://adsabs.harvard.edu/abs/2017ApJ...846...19R} {846, 19}

\bibitem[\protect\citeauthoryear{{Ruge}, {Flock}, {Wolf}, {Dzyurkevich},
  {Fromang}, {Henning}, {Klahr}  \& {Meheut}}{{Ruge}
  et~al.}{2016}]{2016A&A...590A..17R}
{Ruge} J.~P.,  {Flock} M.,  {Wolf} S.,  {Dzyurkevich} N.,  {Fromang} S.,
  {Henning} T.,  {Klahr} H.,   {Meheut} H.,  2016, \mn@doi [\aap]
  {10.1051/0004-6361/201526616}, \href
  {http://adsabs.harvard.edu/abs/2016A%26A...590A..17R} {590, A17}

\bibitem[\protect\citeauthoryear{{Shakura} \& {Sunyaev}}{{Shakura} \&
  {Sunyaev}}{1973}]{1973A&A....24..337S}
{Shakura} N.~I.,  {Sunyaev} R.~A.,  1973, \aap, \href
  {http://adsabs.harvard.edu/abs/1973A%26A....24..337S} {24, 337}

\bibitem[\protect\citeauthoryear{{Shu}}{{Shu}}{1992}]{1992phas.book.....S}
{Shu} F.~H.,  1992, {Physics of Astrophysics, Vol. II}.
University Science Books

\bibitem[\protect\citeauthoryear{{Simon} \& {Armitage}}{{Simon} \&
  {Armitage}}{2014}]{2014ApJ...784...15S}
{Simon} J.~B.,  {Armitage} P.~J.,  2014, \mn@doi [\apj]
  {10.1088/0004-637X/784/1/15}, \href
  {http://adsabs.harvard.edu/abs/2014ApJ...784...15S} {784, 15}

\bibitem[\protect\citeauthoryear{{Simon}, {Pascucci}, {Edwards}, {Feng},
  {Gorti}, {Hollenbach}, {Rigliaco}  \& {Keane}}{{Simon}
  et~al.}{2016}]{2016ApJ...831..169S}
{Simon} M.~N.,  {Pascucci} I.,  {Edwards} S.,  {Feng} W.,  {Gorti} U.,
  {Hollenbach} D.,  {Rigliaco} E.,   {Keane} J.~T.,  2016, \mn@doi [\apj]
  {10.3847/0004-637X/831/2/169}, \href
  {http://adsabs.harvard.edu/abs/2016ApJ...831..169S} {831, 169}

\bibitem[\protect\citeauthoryear{{Squire} \& {Hopkins}}{{Squire} \&
  {Hopkins}}{2017}]{2017arXiv171103975S}
{Squire} J.,  {Hopkins} P.~F.,  2017, preprint, \href
  {http://adsabs.harvard.edu/abs/2017arXiv171103975S} {} (\mn@eprint {arXiv}
  {1711.03975})

\bibitem[\protect\citeauthoryear{{Stepanovs} \& {Fendt}}{{Stepanovs} \&
  {Fendt}}{2014}]{2014ApJ...793...31S}
{Stepanovs} D.,  {Fendt} C.,  2014, \mn@doi [\apj]
  {10.1088/0004-637X/793/1/31}, \href
  {http://adsabs.harvard.edu/abs/2014ApJ...793...31S} {793, 31}

\bibitem[\protect\citeauthoryear{{Stone} \& {Norman}}{{Stone} \&
  {Norman}}{1992a}]{1992ApJS...80..753S}
{Stone} J.~M.,  {Norman} M.~L.,  1992a, \mn@doi [\apjs] {10.1086/191680}, \href
  {http://adsabs.harvard.edu/abs/1992ApJS...80..753S} {80, 753}

\bibitem[\protect\citeauthoryear{{Stone} \& {Norman}}{{Stone} \&
  {Norman}}{1992b}]{1992ApJS...80..791S}
{Stone} J.~M.,  {Norman} M.~L.,  1992b, \mn@doi [\apjs] {10.1086/191681}, \href
  {http://adsabs.harvard.edu/abs/1992ApJS...80..791S} {80, 791}

\bibitem[\protect\citeauthoryear{{Suriano}, {Li}, {Krasnopolsky}  \&
  {Shang}}{{Suriano} et~al.}{2017}]{2017MNRAS.468.3850S}
{Suriano} S.~S.,  {Li} Z.-Y.,  {Krasnopolsky} R.,   {Shang} H.,  2017, \mn@doi
  [\mnras] {10.1093/mnras/stx735}, \href
  {http://adsabs.harvard.edu/abs/2017MNRAS.468.3850S} {468, 3850}

\bibitem[\protect\citeauthoryear{{Suzuki}, {Muto}  \& {Inutsuka}}{{Suzuki}
  et~al.}{2010}]{2010ApJ...718.1289S}
{Suzuki} T.~K.,  {Muto} T.,   {Inutsuka} S.-i.,  2010, \mn@doi [\apj]
  {10.1088/0004-637X/718/2/1289}, \href
  {http://adsabs.harvard.edu/abs/2010ApJ...718.1289S} {718, 1289}

\bibitem[\protect\citeauthoryear{{Tabone} et~al.,}{{Tabone}
  et~al.}{2017}]{2017A&A...607L...6T}
{Tabone} B.,  et~al., 2017, \mn@doi [\aap] {10.1051/0004-6361/201731691}, \href
  {http://adsabs.harvard.edu/abs/2017A%26A...607L...6T} {607, L6}

\bibitem[\protect\citeauthoryear{{Takahashi} \& {Inutsuka}}{{Takahashi} \&
  {Inutsuka}}{2014}]{2014ApJ...794...55T}
{Takahashi} S.~Z.,  {Inutsuka} S.-i.,  2014, \mn@doi [\apj]
  {10.1088/0004-637X/794/1/55}, \href
  {http://adsabs.harvard.edu/abs/2014ApJ...794...55T} {794, 55}

\bibitem[\protect\citeauthoryear{{Troland} \& {Crutcher}}{{Troland} \&
  {Crutcher}}{2008}]{2008ApJ...680..457T}
{Troland} T.~H.,  {Crutcher} R.~M.,  2008, \mn@doi [\apj] {10.1086/587546},
  \href {http://adsabs.harvard.edu/abs/2008ApJ...680..457T} {680, 457}

\bibitem[\protect\citeauthoryear{{Turner}, {Fromang}, {Gammie}, {Klahr},
  {Lesur}, {Wardle}  \& {Bai}}{{Turner} et~al.}{2014}]{2014prpl.conf..411T}
{Turner} N.~J.,  {Fromang} S.,  {Gammie} C.,  {Klahr} H.,  {Lesur} G.,
  {Wardle} M.,   {Bai} X.-N.,  2014, \mn@doi [Protostars and Planets VI]
  {10.2458/azu_uapress_9780816531240-ch018}, \href
  {http://adsabs.harvard.edu/abs/2014prpl.conf..411T} {pp 411--432}

\bibitem[\protect\citeauthoryear{{Umebayashi} \& {Nakano}}{{Umebayashi} \&
  {Nakano}}{1981}]{1981PASJ...33..617U}
{Umebayashi} T.,  {Nakano} T.,  1981, \pasj, \href
  {http://adsabs.harvard.edu/abs/1981PASJ...33..617U} {33, 617}

\bibitem[\protect\citeauthoryear{{Wardle}}{{Wardle}}{2007}]{2007Ap&SS.311...35W}
{Wardle} M.,  2007, \mn@doi [\apss] {10.1007/s10509-007-9575-8}, \href
  {http://adsabs.harvard.edu/abs/2007Ap%26SS.311...35W} {311, 35}

\bibitem[\protect\citeauthoryear{{Weidenschilling}}{{Weidenschilling}}{1977}]{1977MNRAS.180...57W}
{Weidenschilling} S.~J.,  1977, \mn@doi [\mnras] {10.1093/mnras/180.1.57},
  \href {http://adsabs.harvard.edu/abs/1977MNRAS.180...57W} {180, 57}

\bibitem[\protect\citeauthoryear{{Whipple}}{{Whipple}}{1972}]{1972fpp..conf..211W}
{Whipple} F.~L.,  1972, in {Elvius} A.,  ed., From Plasma to Planet. p.~211

\bibitem[\protect\citeauthoryear{{Yen}, {Koch}, {Takakuwa}, {Krasnopolsky},
  {Ohashi}  \& {Aso}}{{Yen} et~al.}{2017}]{2017ApJ...834..178Y}
{Yen} H.-W.,  {Koch} P.~M.,  {Takakuwa} S.,  {Krasnopolsky} R.,  {Ohashi} N.,
  {Aso} Y.,  2017, \mn@doi [\apj] {10.3847/1538-4357/834/2/178}, \href
  {http://adsabs.harvard.edu/abs/2017ApJ...834..178Y} {834, 178}

\bibitem[\protect\citeauthoryear{{Youdin} \& {Goodman}}{{Youdin} \&
  {Goodman}}{2005}]{2005ApJ...620..459Y}
{Youdin} A.~N.,  {Goodman} J.,  2005, \mn@doi [\apj] {10.1086/426895}, \href
  {http://adsabs.harvard.edu/abs/2005ApJ...620..459Y} {620, 459}

\bibitem[\protect\citeauthoryear{{Zanni}, {Ferrari}, {Rosner}, {Bodo}  \&
  {Massaglia}}{{Zanni} et~al.}{2007}]{2007A&A...469..811Z}
{Zanni} C.,  {Ferrari} A.,  {Rosner} R.,  {Bodo} G.,   {Massaglia} S.,  2007,
  \mn@doi [\aap] {10.1051/0004-6361:20066400}, \href
  {http://adsabs.harvard.edu/abs/2007A%26A...469..811Z} {469, 811}

\bibitem[\protect\citeauthoryear{{Zhang}, {Blake}  \& {Bergin}}{{Zhang}
  et~al.}{2015}]{2015ApJ...806L...7Z}
{Zhang} K.,  {Blake} G.~A.,   {Bergin} E.~A.,  2015, \mn@doi [\apjl]
  {10.1088/2041-8205/806/1/L7}, \href
  {http://adsabs.harvard.edu/abs/2015ApJ...806L...7Z} {806, L7}

\bibitem[\protect\citeauthoryear{{Zhang}, {Bergin}, {Blake}, {Cleeves},
  {Hogerheijde}, {Salinas}  \& {Schwarz}}{{Zhang}
  et~al.}{2016}]{2016ApJ...818L..16Z}
{Zhang} K.,  {Bergin} E.~A.,  {Blake} G.~A.,  {Cleeves} L.~I.,  {Hogerheijde}
  M.,  {Salinas} V.,   {Schwarz} K.~R.,  2016, \mn@doi [\apjl]
  {10.3847/2041-8205/818/1/L16}, \href
  {http://adsabs.harvard.edu/abs/2016ApJ...818L..16Z} {818, L16}

\bibitem[\protect\citeauthoryear{{Zhu} \& {Stone}}{{Zhu} \&
  {Stone}}{2017}]{2017arXiv170104627Z}
{Zhu} Z.,  {Stone} J.~M.,  2017, preprint, \href
  {http://adsabs.harvard.edu/abs/2017arXiv170104627Z} {} (\mn@eprint {arXiv}
  {1701.04627})

\bibitem[\protect\citeauthoryear{{Zweibel} \& {Yamada}}{{Zweibel} \&
  {Yamada}}{2009}]{2009ARA&A..47..291Z}
{Zweibel} E.~G.,  {Yamada} M.,  2009, \mn@doi [\araa]
  {10.1146/annurev-astro-082708-101726}, \href
  {http://adsabs.harvard.edu/abs/2009ARA%26A..47..291Z} {47, 291}

\bibitem[\protect\citeauthoryear{{van der Plas} et~al.,}{{van der Plas}
  et~al.}{2017}]{2017A&A...597A..32V}
{van der Plas} G.,  et~al., 2017, \mn@doi [\aap] {10.1051/0004-6361/201629523},
  \href {http://adsabs.harvard.edu/abs/2017A%26A...597A..32V} {597, A32}

\makeatother
\end{thebibliography}
\bibliographystyle{mnras}

\bsp	
\label{lastpage}
\end{document}